\documentclass{article}

\usepackage{xcolor,colortbl}
\usepackage{graphicx}
\usepackage{amsmath}
\usepackage{amssymb}
\usepackage{booktabs}
\usepackage{amsthm}
\usepackage{comment}
\usepackage{cancel}
\usepackage[margin=1in]{geometry}
\usepackage{hyperref}
\usepackage{wrapfig}
\usepackage{cleveref}
\usepackage{footnote}
\makesavenoteenv{tabular}

\newtheorem{theorem}{Theorem}

\newtheorem{proposition}[theorem]{Proposition}

\newcommand{\bemc}[1]{{\color{black}{#1}}}
\newcommand{\bacca}[1]{{\textcolor{black}{#1}}}
\newcommand{\julian}[1]{{\textcolor{black}{#1}}}
\newcommand{\y}{\boldsymbol{y}}

\newcommand{\yo}{\y_1}
\newcommand{\yt}{\y_2}

\newcommand{\loss}[1]{\mathcal{L}_{\mathrm{#1}}}
\newcommand{\yp}{\y_1}
\newcommand{\ym}{\y_2}
\newcommand{\const}{\mathrm{const}}
\newcommand{\x}{\boldsymbol{x}}
\newcommand{\z}{\boldsymbol{z}}
\DeclareMathOperator*{\argmin}{arg\,min}
\newcommand{\der}[2]{\frac{\partial #1}{\partial #2}}
\newcommand{\nder}[3]{\frac{\partial^{#3} #1}{\partial {#2}^{#3}}}
\newcommand{\A}{\boldsymbol{A}}

\newcommand{\Id}{\boldsymbol{I}}

\begin{document}

\begin{center}
    \section*{Generalized Recorrupted-to-Recorrupted: \\Self-Supervised Learning Beyond Gaussian Noise}
\end{center}

\begin{center}
    Brayan Monroy$^1$, Jorge Bacca$^1$, and Julian Tachella$^2$\\
    $^1$Universidad Industrial de Santander, Colombia \\
    $^2$CNRS, ENS de Lyon
\end{center}

\begin{center}
    \url{https://github.com/bemc22/GeneralizedR2R}
\end{center}

\begin{abstract}
Recorrupted-to-Recorrupted (R2R) has emerged as a methodology for training deep networks for image restoration in a self-supervised manner from noisy measurement data alone, demonstrating equivalence in expectation to the supervised squared loss in the case of Gaussian noise. However, its effectiveness with non-Gaussian noise remains unexplored. In this paper, we propose Generalized R2R (GR2R), extending the R2R framework to handle a broader class of noise distribution as additive noise like log-Rayleigh and address the natural exponential family including Poisson and Gamma noise distributions, which play a key role in many applications including low-photon imaging and synthetic aperture radar. We show that the GR2R loss is an unbiased estimator of the supervised loss and that the popular Stein's unbiased risk estimator can be seen as a special case. A series of experiments with Gaussian, Poisson, and Gamma noise validate GR2R's performance, showing its effectiveness compared to other self-supervised methods.

\end{abstract}
\section{Introduction}
\label{sec:intro}

Image restoration is essential in many scientific and engineering applications, from medical imaging to computational photography. State-of-the-art image restoration approaches train a deep network to predict clean images from noisy measurements.
However, most approaches are based on supervised learning, which requires clean image datasets for effective deep network training~\cite{izadi2023image}. This reliance introduces significant challenges, such as the scarcity of clean image data in many important medical and scientific imaging applications~\cite{belthangady2019applications}, and the risk of models overfitting or memorizing specific examples~\cite{somepalli2023diffusion, jagielski2022measuring}. As a result, there is growing interest in developing robust self-supervised learning strategies that can operate exclusively on noisy data~\cite{daras2024ambient, lehtinen2018noise2noise, pang2021recorrupted}, representing a promising direction for the future of image processing and computer vision.

Self-supervised image denoising has attracted significant attention for its potential to perform denoising without relying on clean, paired data, making it a baseline for many image restoration techniques. Methods in this domain are often classified based on the level of prior knowledge they require about the noise distribution~\cite{tachella2024unsure}. 

The first class of methods relies on two independent noisy realizations per image to construct a loss that uses these pairs as inputs and targets. Noise2Noise~\cite{lehtinen2018noise2noise} can obtain performance close to supervised learning, but obtaining independent noisy pairs is impossible in many applications. The second class only assumes that the noise distribution is independent across pixels.  Blind Spot Networks, for example, rely on a masking strategy applied to the input or within the network itself \cite{krull2019noise2void, batson2019noise2self, wang2022blind2unblind}. This approach forces the network to predict denoised values based solely on neighboring pixels, leveraging the assumption of local spatial correlations within the image. As the central pixel is not used, these methods generally have suboptimal performance.

The third family of methods assumes that the noise distribution is fully known, and can obtain performances that are on par with supervised learning. Stein's Unbiased Risk Estimator (SURE)~\cite{stein1981estimation} loss matches the supervised loss in expectation, and is available for Gaussian, Poisson~\cite{le2014unbiased} and Poisson-Gaussian noise distributions~\cite{le2014unbiased}. However, SURE requires approximating the divergence of the restoration function~\cite{ramani2008monte}, which can lead to suboptimal results. The computation of the divergence is avoided by Noise2Score~\cite{kim2022noise}, which approximates the score of the noisy distribution during training and leverages Tweedie's formula to denoise at the test time. 
Finally, Recorrupted-to-Recorrupted~\cite{pang2021recorrupted,oliveira2021unbiased} synthetically recorrupts a noisy image into noisy pairs, removing the need to approximate the divergence term, while remaining an unbiased estimator of the supervised loss. However, this method has been mostly demonstrated with Gaussian noise.

This work extends the R2R framework to address noise types beyond the traditional additive Gaussian model by introducing Generalized R2R (GR2R). Our formulation covers two noise modalities: additive noise and noise distributions within the natural exponential family. By generating independent noisy image pairs from a single noisy image following a specified noise distribution, we establish equivalence to supervised learning through traditional mean-square-error (MSE) loss.
Additionally, we demonstrate that the GR2R approach can be seen as a generalization of the well-known SURE loss, which does not require the computation of divergences.
We further propose using the negative log-likelihood loss that takes into account the noise distribution of the recorrupted measurements and show how the method can be adapted to tackle general inverse problems. The GR2R framework enables training any deep network for NEF noise distributions without requiring modifications to the network structure.  We conduct simulations for Gaussian, Poisson, and Gamma noise distributions on image datasets where these distributions are particularly relevant, including magnetic resonance imaging (MRI), natural image denoising, and synthetic aperture radar (SAR). Results demonstrate that GR2R effectively handles diverse noise distributions, achieving a performance on par with fully supervised learning.

\begin{figure*}[!h]
    \centering
    \includegraphics[width=0.9\linewidth]{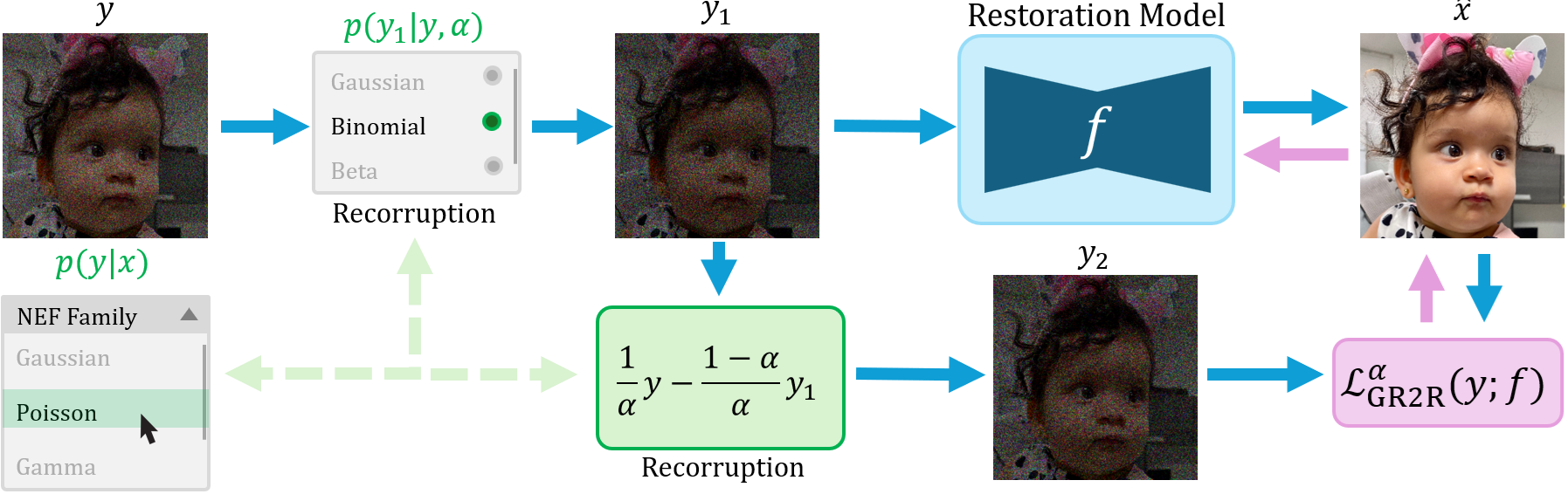} \vspace{-1em}
    \caption{\textbf{Overview of GR2R framework.} Given a noisy image $\y \sim p(\y|\x)$, which belongs to a NEF, GR2R generates pairs of independent noisy images $\y_1$ and $\y_2$ through additional recorruption to enable self-supervised denoising learning $\hat{f}(\y_1) \approx \mathbb{E}\{\x | \y_1\}$.}
    \label{fig:gr2r} \vspace{-1.5em}
\end{figure*}

\section{Related Work}
\label{sec:relatedwork}


\paragraph{Supervised and Self-Supervised Learning.}

The goal of supervised learning is to learn a deep denoiser operator $\hat{\x} = \hat{f}(\y)$ given noisy/clean paired data $(\y, \x)$.  This process can be mathematically descibed by the minimization of the supervised loss function as follows \begin{equation}
   \hat{f}=\argmin_f \mathbb{E}_{\x,\y} \, \loss{SUP}(\x,\y; f), \label{eqn:sup}
\end{equation} where $\loss{SUP}(\x,\y; f) = \Vert f(\y) - \x  \Vert_2^2,$ which is the MSE between the estimation and clean images\footnote{It is important to highlight that other cost functions can be used, such as the mean absolute error.}. The optimal estimator is the Minimum MSE $\hat{f}(\y)\approx \mathbb{E}
\{\x|\y\}$. In practice, the expectation is approximated with a finite dataset with $N$ samples, i.e, $\frac{1}{N}\sum_{i=1}^N \loss{SUP}(\x^{(i)},\y^{(i)};f)$. By simple algebra manipulation $\loss{SUP}$ can be split between unsupervised and supervised terms as follows 
\begin{equation*}
\loss{SUP}(\x,\y; f) = \Vert f(\y) -  \y  \Vert_2^2 + 2 f(\y)^\top( \y - \x) +  \const.
\end{equation*} 
When there is no access to the set of clean images $\x$, self-supervised methods aim to approximate/eliminate the term that contains access to $\x$,  $ f(\y)^\top(\y-\x )$ by building a self-supervised loss $\loss{SELF}$ such that \begin{equation}
    \mathbb{E}_{\y|\x} \loss{SELF}(\y; f) = \mathbb{E}_{\y|\x}\loss{SUP}(\x,\y; f) +  \const. 
\end{equation}


\paragraph{Stein's Unbiased Risk Estimator.}

Assuming that the measurements are corrupted by Gaussian noise, $\y|\x \sim \mathcal{N}(\x,\boldsymbol{I}\sigma^2)$ and $f$ be weakly differentiable,  SURE propose the following self-supervised loss
\begin{equation}
    \loss{SURE}(\y;f)=\Vert f(\y) - \y \Vert_2^2 + 2\sigma^2  \sum_{i=1}^{n} \frac{\partial f_i}{\partial y_i}(\y),
\end{equation}
where the last term is the divergence of $f$ at $\y$. Consequently, SURE guarantees that
\begin{equation*}
    \mathbb{E}_{\y|\x} \{ f(\y)^\top( \y - \x) \} = \mathbb{E}_{\y|\x} \left\{ \sigma^2 \sum_{i=1}^{n} \frac{\partial f_i}{\partial y_i} (\y)\right\},
\end{equation*} therefore, the SURE expectation loss serves as an unbiased risk estimator of its supervised counterpart, with $\hat{f}(\y)\approx \mathbb{E}
\{\x|\y\}$. Noteworthy advancements beyond Gaussian noise include the Hudson lemma~\cite{hudson1978natural}, which extends SURE to the exponential family, and GSURE~\cite{eldar2008generalized}, which further extends it to non-i.i.d exponential families.


\paragraph{Noise2Void and Blind Spot Networks.}

Noise2Void methods and blind spot networks build constrained denoisers $f$ that do not look at the central pixel for denoising, ie $\frac{\partial f_i}{\partial y_i} (\y)= 0$ for all $\y \in \mathbb{R}^n$ and $i=1,\dots,n$. Then, a constrained network $f$ can be trained on simple measurement consistency 
\begin{equation}
    \loss{MC}(\y;f) = \Vert f(\y) - \y  \Vert_2^2,
\end{equation}
where, if the noise distribution is independent across pixels, i.e., $p(\y|\x)=\prod_{i=1}^n p(y_i|x_i)$, then we have that 
\begin{equation*}
    \mathbb{E}_{\y|\x} \{ f(\y)^\top( \y - \x) \} = 0. 
\end{equation*}
Alternatively, approaches such as Neigh2Neigh~\cite{yaman2020self, huang2021neighbor2neighbor} design a custom loss function that removes the central pixel instead of using a blind spot denoiser. However, note that the learned network is not optimal for MMSE because $\hat{f}(\y) \neq \mathbb{E}\{\x|\y\}$, due to the zero derivative constraint.


\paragraph{Noise2Noise.}  Assume a training set of paired noisy/noisy images is available. Specifically, the authors in~\cite{lehtinen2018noise2noise} described the pairs of noisy images as \begin{equation}
    \begin{gathered}
        \y_1 \sim  p(\y_1|\x), \quad  \y_2 \sim   p(\y_2|\x),
     \end{gathered}
\end{equation} then, a denoiser operator $f$ is adjusted to minimize the squared $\ell_2$ loss \begin{equation}
    \loss{N2N}(\y_1,\y_2;f) = \Vert f(\y_1) - \y_2  \Vert_2^2 , \label{eqn:n2n}
\end{equation} since the noisy pairs are independent with $\y_1, \y_2$ independent conditioned on $\x$ and $\mathbb{E}\{\y_1|\x\} = \mathbb{E}\{\y_2|\x\}  = \x$, then the supervised term is reduced such that
\begin{align*}
    &\mathbb{E}_{\y_1, \y_2|\x} \left\{ f(\y_1)^\top(\y_2 - \x) \right\}  \\&=  \mathbb{E}_{\y_1|\x}  f(\y_1)^\top \mathbb{E}_{\y_2| \x} (\y_2 - \x)  = 0,
\end{align*} hence, the Noise2Noise expectation loss closely matches its supervised counterpart $\mathbb{E}_{\y_1,\y_2|\x} \loss{N2N}(\y_1,\y_2;f)  = \mathbb{E}_{\y_1|\x}  \loss{SUP}(\x,\y_1;f) + \const $,  where $\hat{f}(\y_1) \approx \mathbb{E}\{\x|\y_1\}$. However, while the minimization of $\loss{N2N}$ does not require access to clean data, sampling independent pairs of noisy images $(\y_1, \y_2)$ is impractical in many real applications. 


\paragraph{Recorrupted2Recorrupted.} A more realistic scenario is the unsupervised case, with only access to unpaired noisy measurements.  Assuming that the noise distribution is Gaussian $\y \sim \mathcal{N}(0, \Id\sigma^2)$, R2R~\cite{pang2021recorrupted} proposed the re-corruption of noisy measurements as follows
\begin{equation} \label{r2req}
    \y_1 = \y + \tau \boldsymbol{\omega}, \quad \y_2 = \y - \boldsymbol{\omega} / \tau,
\end{equation} 
where $\boldsymbol{\omega} \sim \mathcal{N}(\boldsymbol{0}, \sigma^2 \Id )$. Then, the unsupervised denoising loss is defined as
\begin{equation}
    \loss{R2R} (\y;f) = \mathbb{E}_{\y_1,\y_2|\y} \Vert  f(\y_1) - \y_2 \Vert_2^2.
\end{equation} The new measurements $\y_1$ and $\y_2$ are generated given an single $\y$. While the loss is defined with an expectation, in practice, only a single pair of $(\y_1,\y_2)$ is generated per batch of $\y$.
R2R is equivalent to the supervised cost function defined in Equation~\eqref{eqn:sup} on the re-corrupted pair $\y_1$ since $
 \mathbb{E}_{\y | \x} \, \loss{R2R}(\y;f) =  \mathbb{E}_{\y_1 | \x} \loss{SUP}(\x,\y_1;f) +  \const ,  \label{eq:conclusion_R2R}$ where  $\hat{f}(\y_1)\approx\mathbb{E} \{\x|\y_1\}$. Two recent works have proposed an extension of R2R to Poisson noise~\cite{oliveira2022unbiased,krull2024image} which is similar to the one presented here, while we extend these results to a much larger family of noise distributions, and show its close links with SURE.


\begin{table*}[!t]
\footnotesize 
\centering
\resizebox{\linewidth}{!} {
\setlength\tabcolsep{0.009cm}
\begin{tabular}{c|c|c|c|c}
\toprule
\cellcolor{green!10}\textbf{Model}  & \cellcolor{green!10} \textbf{Gaussian} & \cellcolor{green!10} \textbf{Poisson} & \cellcolor{green!10} \textbf{Gamma} & \cellcolor{green!10} \textbf{Binomial}  \\ 
\cellcolor{green!10}&\cellcolor{green!10} $\y \sim \mathcal{N}(\x,\Id\sigma^2)$ 
& \cellcolor{green!10} $ \boldsymbol{z}\sim \mathcal{P}(\x/\gamma) , \y = \gamma \boldsymbol{z}$ & \cellcolor{green!10} $\y\sim \mathcal{G}(\ell, \ell / \x) $
& \cellcolor{green!10} 
$\boldsymbol{z}\sim \text{Bin}(\ell, \x), \y = \boldsymbol{z} / \ell $
\\ \toprule
$\y_1$ & $ \y_1=\y + \sqrt{\frac{\alpha}{1-\alpha}} \boldsymbol{\omega}$, & $\y_1=\frac{\boldsymbol{y}- \gamma \boldsymbol{\omega}}{1-\alpha}  $, &  $\y_1=\y\circ(\boldsymbol{1}-\boldsymbol{\omega}) /(1-\alpha)$, & $\y_1=\frac{\boldsymbol{y}- \boldsymbol{\omega} / \ell }{1-\alpha}$,  \\ 
&  $\boldsymbol{\omega} \sim  \mathcal{N}(0,\Id\sigma^2)$ & $\boldsymbol{\omega} \sim \text{Bin}(\boldsymbol{z}, \alpha)$  & $\boldsymbol{\omega} \sim  \text{Beta} (\ell \alpha, \ell (1-\alpha))$ &  $\boldsymbol{\omega} \sim \text{HypGeo}(\ell, \ell\alpha, \boldsymbol{z})$ 
\\ \hline
$\y_2$ &  \multicolumn{4}{c}{$ \y_2=\frac{1}{\alpha} \y -  \frac{(1-\alpha)}{\alpha}\y_1$}  \textcolor{white}{\Bigg( \Bigg)}
\\ \hline
 $\loss{GR2R-NLL}^{\alpha}$ & 
 $\|f(\y_1)-\y_2\|^2_2$ & $-\gamma \y_2^{\top}\log f(\y_1)+  \boldsymbol{1}^{\top}f(\y_1)$ &  
 $ \log f(\y_1) +  \y_2/f(\y_1)  $ 
 &  
  \begin{tabular}{c} $ - \y_2^{\top}\log f(\y_1) +$ \\ $ (\y_2-\ell)^{\top}\log(1-f(\y_1)) $
  \end{tabular}  
 \\ \hline
$\loss{GR2R-MSE}^{\alpha}$ &  \multicolumn{4}{c}{$\|f(\y_1)-\y_2\|^2_2$}  \textcolor{white}{\Bigg( \Bigg)}
\\ \hline
\cellcolor{blue!10} \begin{tabular}{c}
   SURE =  \\
    $ \lim\limits_{\alpha \to 0}  \loss{GR2R-MSE}^{\alpha}$
\end{tabular} & \cellcolor{blue!10} \begin{tabular}{c}
    $\|f(\y)-\y\|^2_2+ $  \\
     $ 2\sigma^2 \sum\limits_{i=1}^{n} \der{f_i}{y_i}(\y) $
\end{tabular}  & 
\cellcolor{blue!10}\begin{tabular}{c}
    $\|f(\y)-\y\|^2_2 + $  \\
    $2 \sum\limits_{i=1}^{n} y_i (f_i(\y) - f_i(\y-\gamma \boldsymbol{e}_i)) $
\end{tabular} &  
\cellcolor{blue!10}\begin{tabular}{c}
    $\|f(\y)-\y\|^2_2 + $  \\
    $2 \sum\limits_{i=1}^{n} \sum\limits_{k\geq 1} \frac{b(\ell, k)(-y_i)^{k+1}\Gamma(\ell)}{\Gamma(\ell+k)}  \nder{f_i}{y_i}{k} (\y) $
\end{tabular}
&  
\cellcolor{blue!10}\begin{tabular}{c} not available \footnotemark
\end{tabular} 
\\ \bottomrule 
\end{tabular} } 
\caption{\textbf{Summary of GR2R losses.} Popular noise distributions belonging to the natural exponential family and the associated splitting functions with $\alpha\in (0,1)$, negative-log likelihood losses. As $\alpha\to 0$, the proposed GR2R loss is equivalent to a SURE-based loss, consisting of the quadratic measurement consistency and a divergence-like term. For Gamma noise $b(\ell, k) = ( \ell (k - 1 )) / ( k ( \ell +  k - 1 ) ).$
} \label{tab:NEF} \vspace{-2em}
\end{table*}

\footnotetext[2]{In this case, $\ell \cdot \alpha$ should be a positive integer, thus the limit $\alpha \rightarrow 0$ is not available.}

\section{Generalized R2R}

We extend the R2R loss to different noise distributions. In particular, we present extensions to i) additive noise distributions beyond Gaussian noise and ii) noise distributions belonging to the natural exponential family.

\subsection{Non-Gaussian Additive Noise}
 Assuming that $\y=\x+\boldsymbol{\epsilon}$ where the only condition is that $p(\boldsymbol{\epsilon})$ is independent of $\x$ we have the following proposition.
\begin{proposition} \label{prop:additive_R2E}
    Let $\y=\x+\boldsymbol{\epsilon}$ with noise distribution $p(\boldsymbol{\epsilon})$ independent of $\x$ and let the estimator $f$ be analytic. If $\y_1$ and $\y_2$ are sampled according to \eqref{r2req}, then
    \begin{align*}
    &\loss{R2R}(\y;f)  \propto \mathbb{E}_{\boldsymbol{\epsilon}, \boldsymbol{\omega}}\|f(\y + \tau \boldsymbol{\omega}) - \x\|^2_2  \\ 
   & + \sum_{i=1}^{n} \sum_{k=0}^{\infty} \frac{1}{k!}\mathbb{E}_{\boldsymbol{\epsilon}_{-i},\boldsymbol{\omega}_{-i}} \{ \nder{f_i}{x_i}{k}(x_i; \y_{-i}+\tau \boldsymbol{\omega}_{-i})\} \,  \mathbb{E}_{{\epsilon}_i,{\omega}_i} \,  ({\epsilon}_i-\frac{\omega_i}{\tau}) ({\epsilon}_i+\tau \omega_i)^{k}, 
    \end{align*}
    where $\boldsymbol{\omega}$ is drawn independently of $\boldsymbol{\epsilon}$, $\boldsymbol{y}_{-i}=[y_1,\dots,y_{i-1},y_{i+1},\dots,y_{n}]^{\top}$ and the proportionality ignores any term not including $f$.
\end{proposition}

The proof is included in Appendix~\ref{app: proofs}. In the original R2R case, $\boldsymbol{\epsilon}$ and $\boldsymbol{\omega}$ are assumed to follow the same Gaussian distribution, then $(\boldsymbol{\epsilon}-\frac{\boldsymbol{\omega}}{\tau})$ and $(\boldsymbol{\epsilon}+\tau \boldsymbol{\omega})$ are independent random variables, meaning that the error term is zero.
Proposition~\ref{prop:additive_R2E} shows that the R2R loss can still be accurate beyond this assumption, showing that low-order functions, e.g., linear reconstruction $f$, only require that $\boldsymbol{\omega}$ matches the second-order moment ($k=1$) of $\boldsymbol{\epsilon}$, as
\begin{equation} \label{eq: second order}
    \mathbb{E} \, \omega_i^2 = \mathbb{E}\, \epsilon_i^2,
\end{equation}
for $i=1,\dots,n$, whereas for quadratic functions, the case $k=2$ imposes the constraint
\begin{equation} \label{eq: third order}
    \mathbb{E}\, \omega_i^3 = \frac{1}{\tau} \mathbb{E}\, \epsilon_i^3,
\end{equation}
higher-order functions will require matching higher-order moments $k>2$.
\subsection{Beyond Additive Noise: Exponential family}

In many applications, the noise affecting the measurements is not additive. For example, Poisson noise arises in photon-counting devices~\cite{li2021photon}, and the Gamma distribution is used to model speckle noise~\cite{parhad2024speckle}. Our main theorem extends the R2R framework to a much larger family of noise distributions, the NEF.
Specifically, we assume that we obtain a measurement $\y \sim p(\y|\x)$ where the observation model belongs to the NEF whose density is described as 
\begin{equation} \label{eq: nef}
    p(\y|\x)= h(\y) \exp( \y^{\top} \eta(\x) - \phi(\x)),
\end{equation}
where $h$, $\eta$ and $\phi$ are known function according to the specific distribution (see Appendix~\ref{app: additional information} for examples). Among the NEF are the popular Gaussian, Poisson, Gamma, Binomial, and Beta noise distributions.
For this family of measurements distribution, we generalize the corruption strategy as
\begin{align}
    \y_1 \sim & \; p(\y_1| \y, \alpha), \\ 
    \y_2 = &  \frac{1}{\alpha} \y -  \frac{(1-\alpha)}{\alpha}\y_1,
\end{align}
where $p(\y_1| \y, \alpha)$ is the conditional distribution of $\y_1$ given $\y$, then, we generalize the MSE loss as
\begin{equation}
   \loss{GR2R-MSE}^{\alpha}(\y;f)=\mathbb{E}_{\y_1,\y_2|\y,\alpha}  \Vert f(\y_1) - \y_2 \Vert_2^2. \label{eq:Pr2r}
\end{equation}
According to the noise distribution of $\y$, we propose obtaining $\y_1$ and $\y_2$ in different ways, as seen in Table \ref{tab:NEF}.

If the noise is Gaussian, we have that $p(\y_1|\y,\alpha) =\mathcal{N}(\y,\frac{\alpha  \boldsymbol{I}}{1-\alpha})$, and the proposed loss boils down to the standard R2R loss in Eq.~\eqref{r2req} under the change of variables $\tau = \sqrt{\frac{\alpha}{1-\alpha}}$. The following theorem demonstrated that if the pair images are generated according to Table~\ref{tab:NEF}, the proposed GR2R strategy is equivalent to the supervised loss.

\setcounter{theorem}{0}
\begin{theorem} \label{theo: NEF}
    Let $p(\y|\x)$ density function of $\y$ that belong to the NEF (up to a transformation of $\x$); with $\mathbb{E}\,\{\y|\x\}=\x$. The measurements admit the decomposition $\y=(1-\alpha)\yo  +\alpha\yt $ with $\alpha\in[0,1]$ where $\yo$ and $\yt$ are independent random variables, generated according to Table~\ref{tab:NEF}, whose distribution $p_1(\y_1|\x)$ and $p_2(\y_2|\x)$ also belong to the NEF, with means $ \mathbb{E}\{\y_1|\x\}= \mathbb{E}\{\y_2|\x\}= \x$, and variances $\mathbb{V}\{\y_1 | \x\}= \frac{\mathbb{V}\{\y | \x\}}{1-\alpha} $ and $\mathbb{V}\{\y_2 | \x \}= \frac{\mathbb{V}\{\y | \x\}}{\alpha} $. Moreover, the conditional distributions $p(\y_1|\y)$ and $p(\y_2|\y)$ do not depend on $\x$, we have that
\begin{equation*}
\mathbb {E}_{\y|\x} \loss{GR2R-MSE}^{\alpha}(\y;f) = \mathbb {E}_{\y_1|\x} \Vert
 f(\y_1) -\x \Vert_2^2.
\end{equation*}
\end{theorem}
The proof is included in Appendix~\ref{app: proofs}. One important aspect is the parameter $\alpha$, consequently, by defining the signal-to-noise ratio (SNR) of a random variable $\boldsymbol{z}$ as
$
\text{SNR}(\boldsymbol{z}) := \frac{\| \mathbb{E}\,\boldsymbol{z} \|^2_2}{\mathbb{E}\,\|\boldsymbol{z}-\mathbb{E} \boldsymbol{z}\|^2_2} 
$
we have that the SNR of the new variables is
\begin{align*}
\text{SNR}(\y_1) &= (1-\alpha) \text{SNR}(\y), \\
\text{SNR}(\y_2) &= \alpha \, \text{SNR}(\y),
\end{align*}
and thus have a strictly lower SNR than the original measurement $\y$. A good choice of $\alpha$ should strike the right balance between having inputs $\y_1$ with SNR close to that of $\y$, without having too noisy outputs $\y_2$. In Appendix~\ref{app: experiments}, we evaluate different values of $\alpha$ in different noise settings, finding in most cases that the best $\alpha$ lies in the interval $[0.1,0.2]$.

\paragraph{Equivalence with SURE Loss.}

Interestingly, our GR2R using MSE unsupervised loos is a generalization of the SURE loss when $\alpha \to 0$ without the need to calculate the divergence term. To show that, let us introduce the following proposition. 
\begin{proposition}
Assume that $f$ is analytic, $p(\y|\x)$ belongs to the NEF, and that
$a_k:\mathbb{R}\mapsto\mathbb{R}$ as 
    $$
    a_k(y_i) = \lim_{\alpha\to 0} \mathbb{E}_{y_{2,i}|y_i,\alpha}(y_{2,i}-y_i)(\alpha y_{2,i})^{k},
    $$ 
for all $i=1,\dots,n$ verifies $|a_k(y_i)|<\infty$ for all positive integers $k\geq 1$.  Then,
 \begin{align*}
  &\lim_{\alpha\to 0} \loss{GR2R-MSE}^{\alpha}(\y;f) = \\
  &\| f(\y) - \y\|^2_2  +  2 \sum_{i=1}^n \sum_{k\geq 1} (-1)^{k+1} a_k(y_i) \bacca{\frac{1}{k!}}\frac{\partial^k f_i}{\partial y_i^k}(\y) + \const.
 \end{align*}
\end{proposition}
The proof is included in~\julian{Appendix~\ref{app: proofs}}. With these results, \Cref{tab:NEF} shows the equivalence for various popular noise distributions belonging to the natural exponential family. Interesting,  the standard Gaussian case, we have $a_1(y_i)=\sigma^2$ and $a_k(y_i)=0$ for $k\geq 2$, recovering the standard SURE formula.
In general, if the estimator verifies $\frac{\partial^k f_i}{\partial y_i^k}(\y) \approx 0$ for higher order derivatives $k\geq 2$, we have
 \begin{align*}
  \lim_{\alpha\to 0} & \loss{GR2R-MSE}^{\alpha}(\y;f) \approx \\
  &\| f(\y) - \y\|^2_2  +  2 \sum_{i=1}^n a_1(y_i) \frac{\partial f_i}{\partial y_i}(\y) + \const,
 \end{align*}
where $a_1(y_i) =  \lim\limits_{\alpha \to 0} \mathbb{V}\{y_{2,i}|y_i,\alpha\}$.

\paragraph{Connection with GSURE.}
The popular GSURE loss introduced by Eldar~\cite{eldar2008generalized} is defined as 
$$\loss{GSURE}(\y;f) = \| f(\y) + \nabla \log h(\y) \|^2_2 + 2 \sum_{i=1}^n \der{f_i}{y_i}(\y),$$
and is an unbiased estimator of 
\begin{align}
  \mathbb{E}_{\y|\x}\loss{GSURE}(\y;f) =  \mathbb{E}_{\y|\x} \| f(\y) - \eta(\x)\|^2_2,
\end{align} 
 where $\eta$ and $h$ are the functions associated with the exponential family decomposition in Eq.~\eqref{eq: nef}.
 The equivalence of our method compared with GSURE only occurs for the Gaussian noise case (up to scaling of $f(\y)$ by $\sigma^2$), where $h(\y)=-\sigma^2\y$ and $\eta(\x) = \sigma^2 \x $. 
The GSURE loss targets the estimator $\hat{f}(\y) \approx \mathbb{E} \{ \eta(\x) | \y\} $, whereas in this work we focus on the conditional mean\footnote{Note that in general  $\eta^{-1}(\mathbb{E} \{ \eta(\x) | \y\}) \neq \mathbb{E} \{ \x | \y\}$  for an invertible $\eta$, except for the case of linear $\eta$. } $\hat{f}(\y)\approx \mathbb{E}\{\x |\y\} $.

\subsection{Negative-Log Likelihood Loss}
If $\yp$ and $\ym$ are independent, we could use the MSE unsupervised loss, as is detailed in the previous section, as an unbiased estimator of the supervised loss $\mathcal{L}_{\text{SUP}}(\x, \y_1; f)$.  Although the choice of $\y_1$ and $\y_2$ are based on the noise distribution $\y$, the MSE loss is suboptimal since it does not try to maximize the probability likelihood of the $\y$ noise distribution. We propose instead to minimize the negative log-likelihood (NLL) as the loss function 
\begin{gather*}
 \loss{GR2R-NLL}^{\alpha}(\y;f)  = \mathbb{E}_{\y_1,\y_2|\y}  \{ \phi(f(\y_1))  - \y_2^{\top} \eta(f(\y_1))  \} \\
=  \mathbb{E}_{\y_1,\y_2|\y} \{-  \log p_2\left(\ym|\hat{\x}=f(\yp)\right) \} + \text{const},
\end{gather*} 
where $p_2$ corresponds to the distribution of $\y_2$ given $\x$, which belongs to the exponential family, as shown in~\Cref{theo: NEF}. 
For the additive Gaussian noise case, the negative log-likelihood function reduces to the standard MSE, i.e, $-\log p_2\left(\ym|f(\yp)\right) \propto  \| f(\yp) - \ym\|^2_2 $. \Cref{tab:NEF} shows the losses for popular noise distributions belonging to the NEF. This loss is un unbiased estimator of a supervised negative-log-likelihood loss, whose optimal solution is also the condition mean (the proof is included in~\julian{Appendix~\ref{app: proofs}}).
\begin{proposition}
Under the assumptions of \Cref{theo: NEF}, we have that   \begin{align*} 
        \mathbb{E}_{\y | \x} \, \loss{GR2R-NLL}^{\alpha}&(\y;f) = \\  
        \mathbb{E}_{\y_1|\x} &\{ - \log p_2\left(\x|f(\yp)\right) \} + \const,
    \end{align*} 
    whose minimizer is the minimum mean squared error estimator $f(\yp) = \mathbb{E} \{ \x|\yp\}$.
\end{proposition}

\subsection{Test Time Inference}

Once the deep denoiser $\hat{f}$ is trained in a self-supervised way,
the following Monte Carlo approximation is used to mitigate the effect of re-corruption \begin{equation}
    \hat{\x} \approx \frac{1}{J} \sum_{j=1}^J \hat{f}\left( \y_1^{(j)} \right). \label{eqn:montecarlo}
\end{equation} where $\y_1^{(j)}\sim p(\y_1 |\y, \alpha)$ are i.i.d. samples. \\ \vspace{-1em}

\section{Simulations and Results}

In this section, we extend the experimental validation of the GR2R framework to denoising tasks beyond the additive Gaussian noise model, specifically addressing log-Rayleigh, Poisson and Gamma noise. We conduct a comparative analysis with state-of-the-art unsupervised denoising methods, including SURE~\cite{ramani2008monte}, PURE~\cite{le2014unbiased}, Neigh2Neigh~\cite{huang2021neighbor2neighbor}, and Noise2Score~\cite{kim2021noise2score}, to benchmark the performance of GR2R against unbiased estimators, as well as Noise2X paradigms. To further strengthen our analysis, we replicate the original R2R framework's reported performance on Gaussian noise. GR2R is agnostic to the choice of architecture; therefore, the DRUnet~\cite{zhang2021plug} architecture was utilized for Poisson denoising, while DnCNN~\cite{zhang2017beyond} was used for denoising under the Gamma and Gaussian noise model. The code implementation was developed using the DeepInverse~\cite{Tachella_DeepInverse_A_deep_2023} library and is available as open source.

\subsection{Non-Gaussian Additive Noise}

We first evaluate the robustness of the GR2R for non-Gaussian additive noise using the DIV2K dataset with crops of $512 \times 512$ pixel resolution, with the train split (900 images) for training and the validation split (100 images) for testing. The dataset was corrupted with log-Rayleigh noise~\cite{rivet2007log}, which is non-symmetric, normalized to have a standard deviation $\sigma=0.1$ and zero mean. We compare with the standard R2R~\cite{pang2021recorrupted} approach,  which recorrupts the noisy data using Gaussian additive noise until matching the second order moment, the noise variance, i.e., $\boldsymbol{\omega} \sim \mathcal{N}( \boldsymbol{0},\Id \sigma^2$). We train for 100 epochs with a batch size of 16 and an initial learning rate of 5e-4. In our GR2R approach, we generate additional noise  $\boldsymbol{\omega}$ following~\Cref{prop:additive_R2E} by sampling using a maximum entropy approach such that the second and third order moments conditions in Eq.~\eqref{eq: second order} and Eq.~\eqref{eq: third order} hold. Precisely, $\boldsymbol{w}$ is adjusted to match until the third order moment using a gradient descent algorithm, as presented by the authors in~\cite{bruna2013audio}. We computed $J=15$ Monte Carlo Samples for the inference according to Eq.~\eqref{eqn:montecarlo} for both scenarios.

The results are shown in~\Cref{tab: results additive}. Adding the third-order correction significantly improves more than 3 dB compared with the baseline R2R, being only 0.5 dB below the supervised learning setting, with a similar variance of $\pm 1.51$ among estimations. The reader is referred to \julian{Appendix~\ref{app: experiments}} for additional information on the sampling algorithm.

\begin{table}[!h] \label{tab: results additive}  \centering
\caption{Mean and Variance of the PSNR/SSIM scores with the additive log-Rayleigh noise.} 
\begin{tabular}{c|c|c} 
\toprule
 R2R~\cite{pang2021recorrupted}  & GR2R (ours) & Supervised                              \\
 2nd Moment & 3rd Moment & Learning \\ \midrule
\begin{tabular}[c]{@{}l@{}}$25.32\pm 0.79$\\ $0.576\pm0.08$\end{tabular} & \begin{tabular}[c]{@{}l@{}}\underline{$29.47 \pm 1.51$}\\ \underline{$0.813\pm 0.04$}\end{tabular} & \begin{tabular}[c]{@{}l@{}} $\boldsymbol{29.93\pm 1.50}$\\ $\boldsymbol{0.831\pm 0.04}$   \end{tabular} \\ \bottomrule
\end{tabular}
\end{table}

\break

\subsection{Poisson Noise}

We evaluate our method on the Poisson denoising problem with gain $\gamma$ (as defined in~\Cref{tab:NEF}) using the DIV2K dataset. The dataset, comprising high-resolution images of $512 \times 512$ resolution, was corrupted by Poisson-distributed noise. The denoising models were trained with three unsupervised approaches, including PURE and Neigh2Neigh. The training consists of 300 epochs with a batch size of 50 and an initial learning rate of 1e-3, decreasing by a factor of 0.1 during the final 60 epochs. In our GR2R approach, training pairs were generated with a setting of $\alpha = 0.15$, as detailed in the first column of Table~\ref{tab:NEF}. We computed  $J=5$ Monte Carlo samples for the test according to Eq.~\eqref{eqn:montecarlo}.

\begin{figure*}[!t]
    \centering 
    \includegraphics[width=\linewidth]{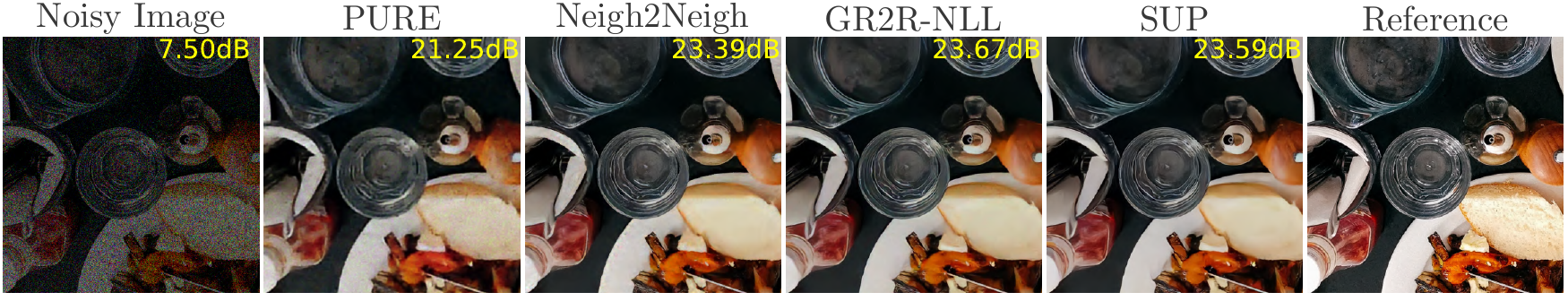}
    \includegraphics[width=\linewidth]{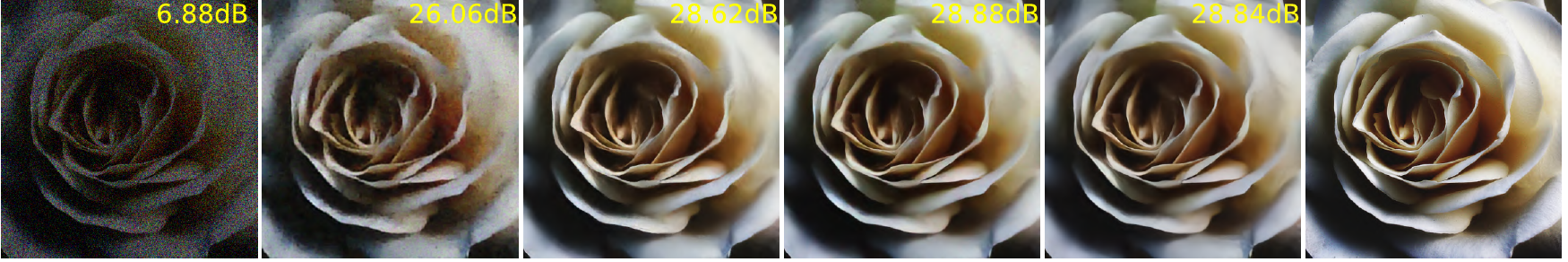}  \vspace{-2em}
    \caption{Denoising results for Poisson noise with $\lambda=0.5$. The first column shows noisy input images, and the last column shows the ground truth reference. Intermediate columns present results from PURE, Neigh2Neigh, GR2R-NLL (proposed), and a supervised MSE-based method (SUP). PSNR values, shown in yellow in the top-right corner of each image, quantify the performance of each method.}
    \label{fig:poisson} \vspace{-1em}
\end{figure*}

Quantitative results for Poisson noise denoising are summarized in Table~\ref{tab:results_poisson}, with visual results presented in Figure~\ref{fig:poisson}. As shown, our proposed GR2R closely match the metrics of the supervised MSE-based model. GR2R-MSE performs well at higher counts (where the noise distribution is closer to Gaussian), while GR2R-NLL loss demonstrates robustness in the low-count regime. Neigh2Neigh is the second-best unsupervised technique but has drawbacks compared to GR2R. It downsamples noisy images, losing high-frequency details, and requires two model evaluations per step, increasing computational costs. GR2R, however, uses a single-term loss and one forward pass per step, making it more efficient and scalable for high-resolution image noise removal.

\begin{table*}[!t]
    \centering  \small
    \caption{PSNR/SSIM results on Poisson noise. GR2R-NLL stands for the proposed GR2R with Negative Log-Likelihood Loss.} 
    \begin{tabular}{l|| cccc|c } \hline 
    Poisson Noise & \multicolumn{4}{c}{Methods}                                                             \\ \hline  
    Noise Level ($\gamma$)  & PURE~\cite{le2014unbiased}            & Neigh2Neigh~\cite{huang2021neighbor2neighbor}        & GR2R-NLL (ours)  & GR2R-MSE (ours)        & Supervised-MSE    \\  \hline \hline
    0.01          &  32.69/0.919             & 33.37/0.929  & 33.90/0.935 & \underline{33.92/0.935} & \textbf{33.96/0.933}  \\ 
    0.1           &  24.37/0.631   & 28.27/0.827  & 28.30/0.827  & \underline{28.35/0.827} & \textbf{28.39/0.827}  \\ 
    0.5           & 22.98/0.623   & 24.90/0.651  & \underline{25.07/0.716}  & 24.69/0.698 & \textbf{25.32/0.727} \\ 
    1.0             & 17.94/0.469   & 23.56/0.653  & \underline{23.69/0.658} &  23.49/0.646 & \textbf{23.85/0.668}  \\ \hline
    \end{tabular} 
    \label{tab:results_poisson} \vspace{-1em}
\end{table*}

\newpage

\subsection{Gamma Noise} 

To examine performance under Gamma noise, we tested on the SAR Image Despeckling dataset with the number of looks $\ell \in \{1, 5, 15, 30\}$~\cite{dalsasso2020sar}.  This dataset includes 127 images with a resolution of $512 \times 512$, each corrupted via Gamma-based sampling to simulate Gamma noise effects. Training under unsupervised methods, including Neigh2Neigh trained for 400 epochs at a batch size of 20. Starting from a learning srate of 1e-3, the rate was decreased by a factor of 0.1 over the last 80 epochs. For the GR2R method, we generate image pairs using the parameters of the first column in Table~\ref{tab:NEF}, with $\alpha = 0.2$. The prediction was performed using Equation~\eqref{eqn:montecarlo}, where $J$ was set to 10.

\begin{figure*}[!h]
    \centering 
    \includegraphics[width=\linewidth]{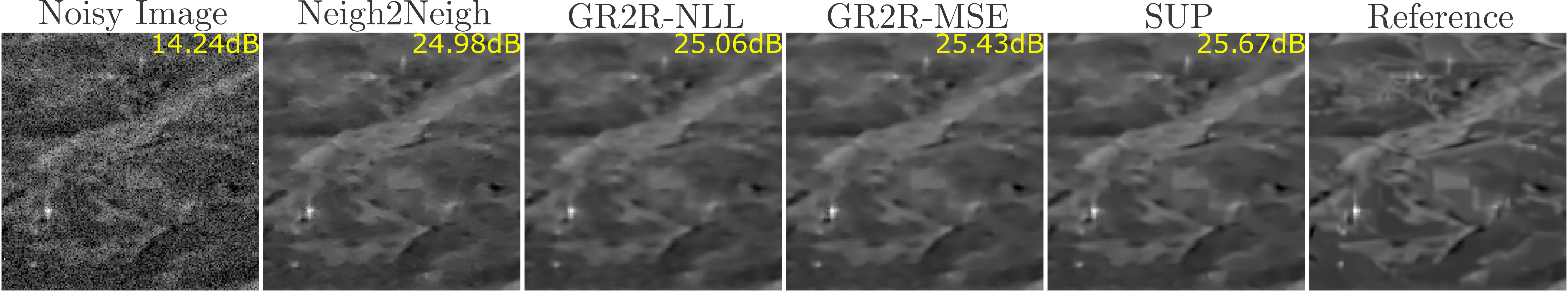}
    \includegraphics[width=\linewidth]{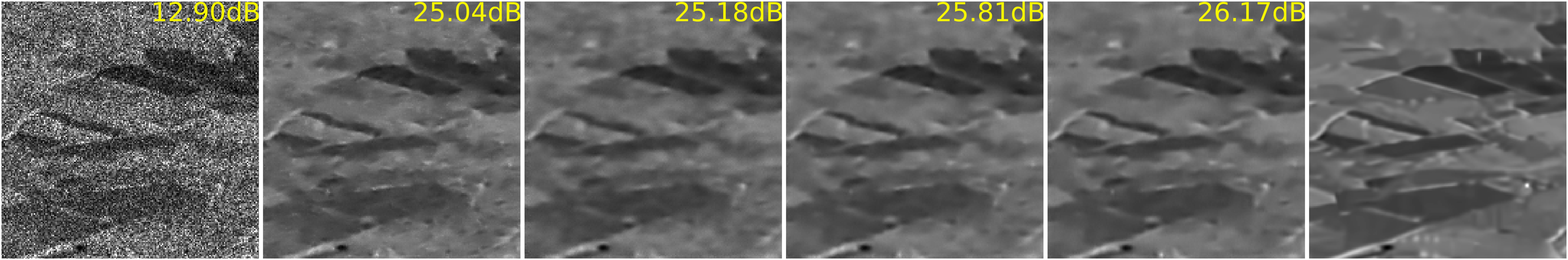}  \vspace{-2em}
    \caption{Denoising results for Gamma Noise with $\ell=5$. The first column shows noisy images, and the last column shows the ground truth reference. Intermediate columns present results from Neigh2Neigh, GR2R-NLL (proposed), GR2R-MSE (proposed) and a supervised MSE-based (SUP). PSNR values, shown in yellow in the top-right corner of each image, quantify the performance of each method.}  \vspace{-1em}
    \label{fig:gamma}
\end{figure*}

Quantitative results for Gamma noise reduction, shown in Table~\ref{tab:results_gamma}, highlight the superior performance of the GR2R framework and its GR2R-MSE variant compared to Neigh2Neigh across all tested noise levels. The supervised MSE-based method achieves the highest PSNR and SSIM, with GR2R-MSE closely matching its performance, particularly at higher SNR ($\ell=30$). The performance between methods converges at lower SNR ($\ell=1$); however, GR2R remains competitive with a PSNR of 32.75 dB, matching Neigh2Neigh and achieving a slightly higher SSIM. Visual comparisons in Figure~\ref{fig:gamma} for Gamma noise at $\ell = 5.0$ demonstrate that GR2R and GR2R-MSE deliver enhanced image quality and reduce noise artifacts more effectively than Neigh2Neigh.

\begin{table*}[!h]
    \centering 
    \caption{PSNR/SSIM results on Gamma noise. GR2R-NLL stands for the proposed GR2R with Negative Log-Likelihood Loss.} 
    \begin{tabular}{l|| ccc|c } \hline
    Gaussian Noise & \multicolumn{4}{c}{Methods}                                                             \\ \hline  
    Number of looks ($\ell$)   & Neigh2Neigh~\cite{huang2021neighbor2neighbor}            & GR2R-NLL (ours)       & GR2R-MSE (ours)         & Supervised-MSE    \\  \hline \hline
    30         & 30.34/0.848   & 30.43/0.862  &  \underline{31.58/0.901} & \textbf{31.86/0.906} \\ 
    15         & 28.56/0.802   &  28.71/0.824  &  \underline{29.55/0.862} & \textbf{29.76/0.865}  \\ 
    5          & 25.71/0.703   &  25.79/0.725  &  \underline{26.35/0.767} & \textbf{26.72/0.784} \\ 
    1          & 22.19/0.560 &   22.19/0.545  &   \underline{22.38/0.599} & \textbf{22.56/0.611}  \\ \hline
    \end{tabular} 
    \label{tab:results_gamma} \vspace{-1em}
\end{table*}

\break

\subsection{Gaussian Noise}


To evaluate the performance of denoising under Gaussian noise, we tested our model on the fastMRI dataset, applying noise levels of $\sigma \in \{0.05, 0.1, 0.2, 0.5\}$. The dataset, comprising 900 images at a resolution of $256 \times 256$, was corrupted with additive Gaussian white noise of zero mean (AGWN) at each level. Training was carried out using four unsupervised techniques: Noise2Score, SURE, Neigh2Neigh, and our proposed GR2R framework, for 600 epochs with a batch size of 32. The learning rate was initialized at 1e-3 and reduced by a factor of 0.1 over the final 48 epochs. GR2R generated training pairs in line with the configuration in the first column of Table~\ref{tab:NEF}, with $\alpha$ set to 0.5. Inference was performed using Equation~\eqref{eqn:montecarlo} with $J = 5$.

Quantitative results, summarized in Table~\ref{tab:results_gaussian}, show that GR2R achieves a denoising performance comparable to the supervised MSE baseline at all noise levels. This outcome is expected, given the equivalence between GR2R training and supervised training, where independent noisy image pairs are generated. Specifically, at $\sigma = 0.05$, GR2R achieves PSNR and SSIM values nearly identical to the highest performing supervised method (35.38 dB vs. 35.41 dB). At elevated noise levels, specifically $\sigma = 0.2$ and $\sigma = 0.5$, GR2R demonstrates robust denoising capabilities (30.24 dB and 25.81 dB), closely aligned with the supervised results (30.38 dB and 25.93 dB) and surpassing other self-supervised approaches.

\begin{table*}[!h]
    \centering
    \caption{PSNR/SSIM results for Gaussian noise. In the case of Gaussian noise for GR2R, the MSE and NLL variants are the same.} 
    \begin{tabular}{l||cccc|c } \hline
    Gaussian noise & \multicolumn{4}{c}{Methods}                                                             \\ \hline  
    Noise Level ($\sigma$) & Noise2Score~\cite{kim2021noise2score}  & SURE~\cite{ramani2008monte}            & Neigh2Neigh~\cite{huang2021neighbor2neighbor}        & GR2R (ours)         & Supervised-MSE    \\  \hline \hline
    0.05         &  34.42/0.916 & 35.31/0.934  &  35.07/0.933  &  \underline{35.38/0.933} & \textbf{35.41/0.935} \\ 
    0.1          &  31.02/0.766 & 32.76/0.910   &  32.57/0.908  &  \underline{33.03/0.911} & \textbf{33.14/0.913}  \\ 
    0.2          &  29.34/0.767 & 29.77/0.874   &  29.73/0.870  &  \underline{30.24/0.880} & \textbf{30.38/0.883} \\ 
    0.5          &  22.94/0.433 & 25.52/0.714 & 25.61/0.717  & \underline{25.81/0.810} &  \textbf{25.93/0.813}  \\ \hline
    \end{tabular}      
    \label{tab:results_gaussian} \vspace{-1em}
\end{table*}

\section{Extension to General Inverse Problems}
\begin{figure}[!h]
    \centering
    \includegraphics[width=\linewidth]{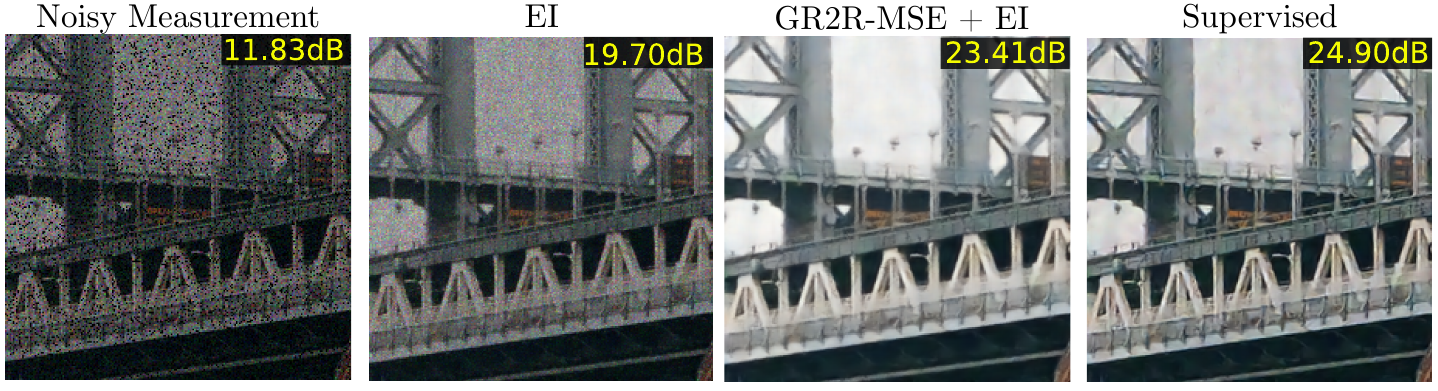} \vspace{-1em}
    \caption{\textbf{GR2R on Inpainting with Poisson Noise.} We compared GR2R with the equivariant imaging framework, consisting of $\mathcal{L}_{\text{MC}} + \mathcal{L}_{\text{EI}}$, and supervised training. The the GR2R framework consists of $\mathcal{L}_{\text{GR2R-MSE}} + \mathcal{L}_{\text{EI}}$. Our findings confirm the effectiveness of GR2R for application in general inverse problems.} 
    \label{fig:inverse}
\end{figure}

The proposed method can be extended to general inverse problems $\y \approx \A\x$ with $\A\in\mathbb{R}^{m\times n}$ where the observation model $p(\y|\A\x)$ belongs to the NEF with mean $\mathbb{E}\{\y | \x\} = \A \x$, by considering the main loss function as
\begin{equation}
\loss{GR2R-MSE}^{\alpha}(\y;f)  =  \mathbb{E}_{\y_1,\y_2|\y} \; \| \A f(\y_1) - \y_2 \|^2_2, 
\end{equation}
which is an unbiased estimator of the clean measurement consistency loss
\begin{equation*}
\mathbb{E}_{\x,\y}  \loss{GR2R-MSE}^{\alpha}(\y;f)  = \mathbb{E}_{\x,\y_1} \;   \|\A ( f(\y_1) -\x) \|^2_2. 
\end{equation*}
This adaptation is straightforward from denoising to the general inverse problem in contrast to other methods, such as Neigh2Neigh, requiring that $\y$ is in the image domain. If  $\text{rank}(\A)=n$, then the GR2R minimizer is the optimal MMSE estimator, that is $\hat{f}(\y_1)\approx\mathbb{E}\{ \x |\y_1\}$. However if $\text{rank}(\A)<n$, then $\A$ has a non-trivial nullspace and $\hat{f}(\y_1) \neq \mathbb{E}\{ \x |\y_1\}$. In this case, it is still possible to learn in the nullspace of $\A$ if we can access a family of different $\{\A_g\}_{g=1}^{G}$ operators~\cite{tachella2022unsupervised,daras2024ambient} or if we can assume that the image distribution $p(\x)$ is approximately invariant to a set of transformations $\{\boldsymbol{T}_g \in \mathbb{R}^{n\times n}\}_{g=1}^{G}$, such as translations or rotations~\cite{chen2021equivariant,chen2022robust}. See~\cite{tachella2022unsupervised} for a detailed discussion of self-supervised learning when $\A$ is incomplete. These additional assumptions can be incorporated via a second loss term: in the case of a single operator $\A$ and invariance, we propose to minimize  
\begin{equation}
\loss{GR2R-MSE}^{\alpha}(\y;f) + \loss{EI}(\y;f)
\end{equation}
where the Equivariant Imaging (EI) loss~\cite{chen2021equivariant} is defined as 
\begin{equation}
    \loss{EI}(\y;f) = \mathbb{E}_g \mathbb{E}_{\y_1|\y} \| f(\A \boldsymbol{T}_g\hat{\x})-\boldsymbol{T}_g \hat{\x}\|_{\bemc{2}}^2
\end{equation}
with $\hat{\x}=f(\y_1)$.  Figure~\ref{fig:inverse} presents the results of comparing the EI framework using the $\mathcal{L}_{\text{MC}}$ loss against EI with our GR2R strategy.  We tested GR2R on the DIV2K dataset, using a binary mask $A$ with entries following a Bernoulli distribution with $p=0.9$ (fixed throughout the dataset to maintain a fixed null-sapce), and corrupting measurements with Poisson noise at~$\gamma=0.05$. The results show that while EI effectively addresses the inpainting problem but still preserves the noise artifacts, in contrast, GR2R successfully tackles both denoising and inpainting simultaneously, delivering results that are comparable to its supervised counterpart. More details and experiments using other distributions are shown in \julian{Appendix\ref{app: experiments}}.

\section{Conclusions}

We present the Generalized Recorrupted-to-Recorrupted (GR2R), which extends the original R2R framework to additive noise and natural exponential family (NEF). We specifically evaluate its performance on Poisson, Gamma, and log-Rayleigh noise in addition to Gaussian noise. The key advantages are simplicity and efficiency: GR2R requires only a simple single-term loss and a single forward pass per training step, and it does not rely on continuous approximations to handle discrete noise distributions (e.g., Poisson or Binomial) used in previous methods (e.g., PURE, Noise2Score).
Moreover, GR2R shows that our method recovers SURE-type loss when $\alpha\to 0$. Furthermore, our re-corruption strategy generates independent noisy pairs directly within the measurement space; naturally extends its applicability beyond image denoising to a wide range of self-supervised inverse problems as inpainting.

\section*{Acknowledgements}
Juli\'an Tachella is supported by the ANR grant UNLIP (ANR-23-CE23-0013), and VIE-UIS supports the UIS contribution under grant 3968.

{
    \small
    \bibliographystyle{IEEEtran}
    \bibliography{main}
}

\break 

\appendix

\vspace{1em}

\section{Proofs} \label{app: proofs}
\subsection*{Proof of Proposition 1}

\begin{proof}
\noindent The R2R loss can be re-expressed as
\begin{equation*}
   \mathbb{E}_{\y_1,\y_2|\y}  \| f(\y_1) - \y_2 \|^2_2 = \mathbb{E}_{\y_1|\y}\| f(\y_1) \|^2_2 + \mathbb{E}_{\y_2|\y} \|\y_2\|^2_2  -   2 \sum_{i=1}^{n}   \mathbb{E}_{\y_1,y_{2,i}|\x} \, y_{2,i}  f_i(\y_1), 
\end{equation*}
where $y_{2,i}\in \mathbb{R}$ denotes the $i$th entry of $\y_2$. If the following equality
\begin{equation}\label{eq:cond_sup}
\mathbb{E}_{\y_1,y_{2,i}|\x} \, y_{2,i}  \, f_i(\y_1) =  x_i \, \mathbb{E}_{\y_1|\x} \, f_i(\y_1),
\end{equation}
holds (below is how to ensure this) for all $i=1,\dots,n$, then 
\begin{align}
   \mathbb{E}_{\y_1,\y_2|\x}  \| f(\y_1) - \y_2 \|^2_2 &=   \mathbb{E}_{\y_1|\x}\| f(\y_1) \|^2_2 + \mathbb{E}_{\y_2|\x} \|\y_2\|^2_2 -   2 \sum_{i=1}^{n} x_i  \mathbb{E}_{\y_1|\x} \,  f_i(\y_1)  \\
   &= \mathbb{E}_{\y_2 |\x} \|f(\y_1)-\x\|^2_2 -  \| \x\|^2_2 + \mathbb{E}_{\y_2|\x}  \| \y_2\|^2_2  \nonumber \\
   &= \mathbb{E}_{\y_2|\x} \|f(\y_1)-\x\|^2_2 + \mathrm{const,} \nonumber
\end{align}
where the second line comes from adding and subtracting $||\x||_2$.

A sufficient (but not necessary) condition for \eqref{eq:cond_sup} to hold is that i) $\y_1$ and $\y_2$ are independent and  ii) $\mathbb{E}_{\y_2|\x} \y_2 = \x$. If this conditions hold, we trivially have $\mathbb{E}_{\y_1,\y_{2,i}|\x} y_{2,i}  f_i(\y_1) =  \left(\mathbb{E}_{\y_{2,i}|\x} y_{2,i}  \right) \left(\mathbb{E}_{\y_1 |\x}f_i(\y_1) \right)= x_i \mathbb{E}_{\y_1 |\x}f_i(\y_1)$ for $i=1,\dots,n$. We will analyze the necessary condition (beyond independence) for the case of additive noise where $\y=\x + \boldsymbol{\epsilon}$ where $\epsilon$ is sampled from a symmetric noise distribution that is independent across pixel entries. We construct pairs $\y_1 = \y + \boldsymbol{\omega} \tau $ and $\y_2 = \y-\boldsymbol{\omega}/\tau$, with $\tau > 0$ and $\boldsymbol{\omega}$ sampled from the same distribution as $\boldsymbol{\epsilon}$. Due to the independence across entries, we will drop the $i$th indices and define the scalar function $f_i(\cdot; \y_{1,-i}): \mathbb{R} \mapsto \mathbb{R} $, such that the left-hand side of \eqref{eq:cond_sup} can be simplified to 
\begin{equation}\label{eq:cond_sup2}
\mathbb{E}_{\epsilon_i,\omega_i} (\underbrace{x_i + \epsilon_i- \omega_i/\tau}_{y_{2,i}})  f_i(\underbrace{x_i+ \epsilon_i+\tau \omega_i}_{y_{1,i}}; \y_{1,-i}) = x_i \, \mathbb{E}_{\epsilon_i,\omega_i}f_i(x_i+ \epsilon_i+\tau \omega_i, \y_{-i}) - \mathbb{E}_{\epsilon_i,\omega_i} \, (\epsilon_i-\frac{\omega_i}{\tau}) f_i(x_i+ \epsilon_i+\tau \omega_i; \y_{1,-i}),
\end{equation}
where $\omega_i$, $x_i$ and $\epsilon_i$ refer to the $i$th entry, and are thus one-dimensional. In this additive case, showing \eqref{eq:cond_sup2} is equivalent to showing that
\begin{equation}\label{eq:cond_sup3}
\mathbb{E}_{\epsilon_i,\omega_i} \, (\epsilon_i-\frac{\omega_i}{\tau})f_i(x_i+\epsilon_i +\tau \omega_i; \y_{1,-i}) = 0.
\end{equation}
Assuming that $f_i$ is analytic (that is, is infinitely differentiable and has a convergent Taylor expansion) and performing a Taylor expansion of $f_i$ around $x_i$, we obtain
\begin{align}
\nonumber
\mathbb{E}_{\boldsymbol{\epsilon}_{-i},\boldsymbol{\omega}_{-i}} \mathbb{E}_{\epsilon_i,\omega_i} \, (\epsilon_i-\frac{\omega_i}{\tau}) f_i(x_i+ \epsilon_i+\omega_i \tau; \y_{1,-i})  &= \mathbb{E}_{\boldsymbol{\epsilon}_{-i},\boldsymbol{\omega}_{-i}} \mathbb{E}_{\epsilon_i,\omega_i} \,   (\epsilon_i-\frac{\omega_i}{\tau}) \sum_{k\geq 0}  \frac{1}{k!}\nder{f_i}{x_i}{k}(x_i; \y_{1,-i}) \, (\epsilon_i+\tau \omega_i)^{k} \\ \nonumber
 &= \sum_{k \geq 0}  \frac{1}{k!} \mathbb{E}_{\boldsymbol{\epsilon}_{-i},\boldsymbol{\omega}_{-i}} \{ \nder{f_i}{x_i}{k}(x_i; \y_{1,-i})\} \,  \mathbb{E}_{\epsilon_i,\omega_i} \,  (\epsilon_i-\frac{\omega_i}{\tau}) (\epsilon_i+\tau \omega_i)^{k}
\end{align}
where the case $k=0$ is removed from the last sum as $\mathbb{E}_{\epsilon_i,\omega_i}\{\epsilon_i-\frac{\omega_i}{\tau}\}= 0$ if the two noises have zero mean. 




\end{proof}

\newpage

\subsection*{Proof of Theorem 1}
\begin{proof}
If the observation model belongs to the natural exponential family (NEF), we can write it as 
$$
p(\y|\x)= h(\y) \exp( \y^{\top}\eta(\x) - \phi(\x)),
$$
with $\x,\y\in\mathbb{R}^n$, and $h:\mathbb{R}\mapsto\mathbb{R}$ $\eta:\mathbb{R}\mapsto\mathbb{R}$ and $\phi:\mathbb{R}\mapsto\mathbb{R}$ elementwise functions which change according to the distribution. NEF distributions verify the following properties~\cite{efron2022exponential}
\begin{enumerate}
    \item  $\eta$ is an invertible function. 
    \item $\phi$ is strictly convex.
    \item Given $\phi$ and $\eta$, $h$ is given by the Laplace transform $h(\y)=\int \exp \left(-\boldsymbol{s}^{\top}\y+\phi(\eta^{-1}(\boldsymbol{s}))\right)d\boldsymbol{s}$.
    \item The mean of each entry is given by 
    \begin{equation} \label{eq: nef mean}
    \mathbb{E} \{y_i|x_i \}= \der{\phi}{x_i}(x_i)/\der{\eta}{x_i}(x_i) = x_i,
    \end{equation}     for $i=1,\dots,n$.
\end{enumerate}
We look for the decomposition $\y= (1-\alpha) \y_1 +  \alpha\y_2$ such that $\y_1$ and $\y_2$ also belong to the NEF, i.e., 
\begin{equation}
p_1(\y_1|\x) = h_1(\y_1) \exp \left( \y_1^{\top}\eta_1(\x) - \phi_1(\x)\right),
\end{equation}
and
\begin{equation}
p_2(\y_2|\x) = h_2(\y_2) \exp \left( \y_2^{\top}\eta_2(\x) - \phi_2(\x)\right),
\end{equation}
for some $\alpha\in (0,1)$. Hence, the element-by-element functions of $\y_1$, $\y_2$ are related to those of $\y$ as  $\phi_1(\x)=(1-\alpha)\phi(\x)$, $\phi_2(\x)=\alpha\phi(\x)$, $\eta_1(\x) =(1-\alpha) \eta(\x)$, $\eta_2(\x) = \alpha \eta(\x)$,  $h_1(\y_1)=\int \exp\left(-\boldsymbol{s}^{\top}\y_1 +(1-\alpha)\phi\left(\eta^{-1}(\frac{\boldsymbol{s}}{1-\alpha})\right)\right)d\boldsymbol{s}$ and $h_2(\y_2)=\int \exp\left(-\boldsymbol{s}^{\top}\y_2+\alpha\phi\left(\eta^{-1}(\frac{\boldsymbol{s}}{\alpha})\right)\right) d\boldsymbol{s}$. 

\noindent We first verify that this choice gives the right distribution for $\y$:
\begin{align*}
p(\y|\x) &= \int  p_1(\y_1|\x)  p_2(\frac{1}{\alpha}\y-\frac{1-\alpha}{\alpha}\y_1|\x) d\y_1 \\
&= \exp\left(\y^{\top}\eta(\x) -\phi(\x)\right)  \int  h_1(\y_1) h_2(\frac{1}{\alpha}\y-\frac{1-\alpha}{\alpha}\y_1)d\y_1,
\end{align*}
where the second line uses the fact that
\begin{align} 
  p_1(\y_1|\x)  p_2(\frac{1}{\alpha}\y-\frac{1-\alpha}{\alpha}\y_1|\x)
&=   h_1(\y_1) \exp\left( (1-\alpha)\y_1^{\top}\eta(\x) - (1-\alpha)\phi(\x)\right)h_2(\frac{1}{\alpha}\y-\frac{1-\alpha}{\alpha}\y_1) \times\\ &\exp\left( \alpha \left(\frac{1}{\alpha}\y-\frac{1-\alpha}{\alpha}\y_1 \right)^{\top} \eta(\x) - \alpha \phi(\x)\right)\\
&= \exp\left(\y^{\top}\eta(\x) -\phi(\x)\right) h_1(\y_1) h_2(\frac{1}{\alpha}\y-\frac{1-\alpha}{\alpha}\y_1) \label{eq: key property}.
\end{align}
We can obtain the conditional distribution of $\y_1$ given $\y$ as
$$
p(\y_1|\y,\x) = \frac{1}{p(\y|\x)}p(\y|\y_1,\x) p_1(\y_1|\x) ,
$$
due to Bayes theorem. 
Using the fact that $p(\y|\y_1,\x)=p_2(\frac{1}{\alpha}\y-\frac{1-\alpha}{\alpha}\y_1|\x)$ we obtain
\begin{equation}
\begin{aligned}
p(\y_1|\y,\x) &= \frac{1}{p(\y|\x)}p_1(\y_1|\x)p_2(\frac{1}{\alpha}\y-\frac{1-\alpha}{\alpha}\y_1|\x)   \\
&= \frac{h_1(\y_1)h_2(\y-\y_1)}{h(\y)}, \label{eqn:p1x}
\end{aligned}
\end{equation}
where we use again \eqref{eq: key property}. Thus we have that $p(\y_1|\y,\x)$ does not depend on the unknown parameter $\x$, that is $p(\y_1|\y,\x)=p(\y_1|\y)$. Consequently, 
since $\y_1$ and $\y_2$ are independent conditional on $\x$ and $ \mathbb{E}_{\y_2|\x}\{\y_2-\x\}= \mathbb{E}\{\y_2|\x\}-\x = \boldsymbol{0}$, we have that
\begin{align*}
    \mathbb{E}_{\y_1,\y_2|\x} \|  f(\y_1)-\y_2\|^2_2 &=  \mathbb{E}_{\y_1|\x} \|  f(\y_1)-\x\|^2_2 + 2 \mathbb{E}_{\y_1,\y_2|\x} \{(f(\y_1)-\x)^\top(\x-\y_2)\} + \mathbb{E}_{\y_2|\x} \| \x - \y_2\|^2_2 \\ 
    &=  \mathbb{E}_{\y_1|\x} \|  f(\y_1)-\x\|^2_2 + 2  \mathbb{E}_{\y_1|x} \{f(\y_1)\}^\top \mathbb{E}_{\y_2|\x}\{\y_2-\x\}- \underbrace{2\mathbb{E}_{\y_2|\x}\{\x^\top (\x-\y_2)\}   + \overbrace{\mathbb{E}_{\y_2|\x} \| \x - \y_2\|^2_2}^{n\mathbb{V}\{\y_2|\x\}}}_{\mathrm{const}} \\
    &= \mathbb{E}_{\y_1|\x} \|  f(\y_1)-\x\|^2_2 + \mathrm{const}.
\end{align*}
\end{proof}

\subsection*{Proof of Proposition 2}


\begin{proof}
We can write the GR2R-MSE loss as
\begin{align}
\loss{GR2R-MSE}^{\alpha}(\y;f) &= \mathbb{E}_{\y_2|\y} \, \| f(\frac{\y-\y_2 \alpha}{1-\alpha}) - \y_2\|^2_2  \\
&= \mathbb{E}_{\y_2|\y}  \, \| f(\frac{\y-\y_2\alpha}{1-\alpha}) - \y - (\y_2-\y)\|^2_2  \\
&=  \mathbb{E}_{\y_2|\y}  \, \| f(\frac{\y-\y_2\alpha}{1-\alpha}) - \y\|^2_2  -\mathbb{E}_{\y_2|\y} \, 2 \sum_{i=1}^n \left(y_{2,i}-y_i\right) f_i(\frac{\y - \y_2\alpha}{1-\alpha}) +\mathrm{const}.
\end{align}
\noindent
Since by assumption $f$ is analytic, we can apply a Taylor expansion to the second term, i.e., $f_i(\frac{\y - \y_2\alpha}{1-\alpha})  = \sum_{k\geq 0}  \frac{1}{k!} \frac{ \partial^k f_i} {\partial y_i^{k}}(\frac{\y-\y_{2,-i}\alpha}{1-\alpha}) \frac{(-1)^{k}\alpha^k}{(1-\alpha)^{k}} y_{i,2}^{k}$,
where $\y_{2,-i}\in\mathbb{R}^n$ has the $i$th entry equal to zero and the rest equal to $\y_2$. Thus we obtain:
$$
\loss{GR2R-MSE}^{\alpha}(\y;f) \propto \mathbb{E}_{\y_2|\y} \Big(\, \| f(\frac{\y-\y_2\alpha}{1-\alpha}) - \y\|^2_2  -  2 \sum_{i=1}^n \sum_{k\geq 1} \frac{1}{k!} \frac{\partial^k f_i}{\partial y_i^{k}}(\frac{\y-\y_{2,-i}\alpha}{1-\alpha})  \frac{(-1)^{k}}{(1-\alpha)^{k}} (y_{2,i}- y_i) (\alpha y_{2,i})^{k} \Big),
$$
where we used the fact that for $k=0$ we have $\mathbb{E}\{y_{2,i}-y_i | y_i\} = 0$. 
Taking the limit $\alpha \to 0$, we obtain\footnote{We have that for $g:\mathbb{R}^n\mapsto\mathbb{R}$, the expectation $\mathbb{E}_{\y_2|\y} g(\alpha \y_2) = g(\boldsymbol{0})$ as $p(\alpha\y_2|\y) \to \delta_{\y_2=\boldsymbol{0}}$ as $\alpha\to 0$.}
\begin{align*} 
\lim_{\alpha \to 0} \loss{GR2R-MSE}^{\alpha}(\y;f) &\propto  \lim_{\alpha \to 0} 
 \mathbb{E}_{\y_2|\y} \, \| f(\frac{\y-\y_2\alpha}{1-\alpha}) - \y\|^2_2 \\&-  2 \sum_{i=1}^n \sum_{k\geq 1} \frac{1}{k!} \mathbb{E}_{\y_{2,-i}|\y} \left\{ \frac{\partial^k f_i}{\partial y_i^{k}}(\frac{\y-\y_{2,-i}\alpha}{1-\alpha})  \right\} \frac{(-1)^{k}}{(1-\alpha)^{k}} \mathbb{E}_{y_{2,i}|y_i} (y_{2,i}- y_i) (\alpha y_{2,i})^{k}\\
 &\propto   \| f(\y) - \y\|^2_2  +  2 \sum_{i=1}^n \sum_{k\geq 1}  (-1)^{k+1} \frac{1}{k!} \frac{\partial^k f_i}{\partial y_i^{k}}(\y) \lim_{\alpha \to 0}  \mathbb{E}\{(y_{2,i}-y_i)(\alpha y_{2,i})^{k}|y_i, \alpha\}
\end{align*}
where the last line uses the fact that $a_k(y_i) = \lim_{\alpha\to 0} \mathbb{E}_{y_{2,i}|y_i, \alpha}\{(y_{2,i}-y_i)(\alpha y_{2,i})^{k}\} $  converges for all positive integer $k$. Replacing the definition of $a_k$ in the previous formula, we obtain the desired result:  $$
  \lim_{\alpha \to 0} \loss{GR2R-MSE}^{\alpha}(\y;f) \propto \| f(\y) - \y\|^2_2  + 
2 \sum_{i=1}^n \sum_{k\geq 1} (-1)^{k+1} a_k(y_i) \frac{1}{k!} \frac{\partial^k f_i}{\partial y_i^k}(\y).
 $$

\end{proof}

\subsection*{Proof of Proposition 3}

\begin{proof}
The $\loss{GR2R}^{\alpha}(\y;f)$ is defined as
\begin{align} 
    \mathbb{E}_{\y_1,\y_2|\x} -  \log p_2\left(\ym|\hat{\x} =f(\y_1)\right) &=  \mathbb{E}_{\y_1,\y_2|\x} \left\{- \alpha \y_2^{\top} \eta\left(f(\y_1)\right) + \alpha \phi\left(f(\y_1)\right) -\log h_2(\y_2) \right\}  \\ \nonumber 
    &=  - \alpha\left(\mathbb{E}_{\y_2|\x} \, \y_2^{\top} \right) \, \mathbb{E}_{\y_1|\x} \, \eta\left(f(\y_1)\right) + \mathbb{E}_{\y_1|\x} \alpha \phi\left(f(\y_1)\right) - \mathbb{E}_{\y_2|\x}\log h_2(\y_2)  \\ \nonumber
    &= - \alpha\x^{\top} \mathbb{E}_{\y_1|\x} \, \eta\left(f(\y_1)\right) + \mathbb{E}_{\y_1|\x} \, \alpha \phi\left(f(\y_1)\right) -  \mathbb{E}_{\y_2|\x}\log h_2(\y_2) \\ \nonumber
    &=- \mathbb{E}_{\y_1|\x} \{\alpha\x^{\top}\eta\left(f(\y_1)\right) - \alpha \phi\left(f(\y_1)\right) +  \log h_2(\y_2)\} + \mathrm{const}  \\ \nonumber
    & =  \mathbb{E}_{\y_1|\x} -  \log p_2\left(\x|\hat{\x} =f(\y_1)\right) + \mathrm{const.} 
\end{align}
We now prove that $\mathbb{E}\{\x|\y_1\} = \arg \min_f \loss{GR2R}^{\alpha}(\y;f) $. We can write this minimization as 
\begin{align}
 \min_f  \;\mathbb{E}_{\x,\y_1} \{ \eta\left(f(\y_1)\right)^{\top} \x -  \phi\left(f(\y_1)\right) \}  &= \min_f  \; \mathbb{E}_{\y_1} \{ \eta\left(f(\y_1)\right)^{\top}\mathbb{E}\{\x|\y_1\} - \phi\left(f(\y_1)\right) \}
   \\
  &=  \mathbb{E}_{\y_1} \{ \min_f \eta\left(f(\y_1)\right)^{\top}\mathbb{E}\{\x|\y_1\} - \phi\left(f(\y_1)\right) \},
\end{align}
where the last equality swaps the integration with the minimization since the minimizer exists for every fixed $\y_1$.
Defining $\z:=f(\y_1)$, we can minimize the term inside the expectation w.r.t.\ to 
\begin{align}
 \arg \min_{\z} \; \mathbb{E}\{\x|\y_{1}\} \, \eta(\z) - \phi\left(\z\right).
\end{align}
The problem is separable across entries, so it can be \begin{align}
\arg \min_{z_i} \; \mathbb{E}\{x_i|y_{1,i}\} \, \eta(z_i) - \phi\left(z_i\right),
\end{align}
for $i=1,\dots,n$. Since the problem is strongly convex w.r.t.\ $z_i$, we can find the solution by setting its derivative to zero
\begin{align}
 \mathbb{E}\{x_i|y_{1,i}\} \, \der{\eta}{z_i}(\hat{z}_i) - \der{\phi}{z_i}\left(\hat{z}_i\right) &= 0 \\
 \der{\eta}{z_i}(\hat{z}_i) /\der{\phi}{z_i}\left(\hat{z}_i\right) &=  \mathbb{E}\{x_i|y_{1,i}\} \\
 \hat{z}_i &= \mathbb{E}\{x_i|y_{1,i}\},
\end{align}
for $i=1,\dots,n$, where the second line uses property \eqref{eq: nef mean}, and thus $\hat{f}(\y_1) = \mathbb{E}\{\x|\y_{1}\}$.

\end{proof}

\section{Additional information} \label{app: additional information}
Table~\ref{table:distributions} summarizes the NEF distributions $p(\y|\x)$ used in the main document. This was used to create the recoruptions used in the main document. Specifically, the formulas to construct $\y_1$ in terms of $\y$ and the extra noise $\boldsymbol{\omega}$ can be derived from replacing $h(y ), h_1(y_1)$ and $h_2(y_2)$ from Table~\ref{table:distributions} in Equation~\eqref{eqn:p1x} for its respective NEF distribution; this is left as an exercise for the reader.
\begin{table}[ht]
\footnotesize
\centering
\begin{tabular}{l|l|l|l|l}
\hline
Model & \begin{tabular}[c]{@{}l@{}}$y\sim\mathcal{N}(x,\sigma^2)$ \end{tabular} & $y\sim \mathcal{P}(\frac{x}{\gamma})$ & $y\sim \mathcal{G}(\ell, x/\ell)$ & $y\sim \text{Bin}(\ell, x)$ \\ \hline
$\eta(x)$ & $x/\sigma^2$ & $\log(x) $ & $-\ell/x$ &  $\log (x/(1-x))$ \\ \hline
$\phi(x)$ & $x^2/(2\sigma^2)$ & $x/\gamma$ & $\ell \log (x)$& $\ell\log(1-x)$ \\ \hline
$h(y)$ & $\sqrt{2\pi}\sigma\exp(y^2/(2\sigma^2))$& $(\gamma^{y}y!)^{-1}$ & $\ell^{\ell}y^{\ell-1}/\Gamma(\ell)$ & $ \ell \choose y$\\ \hline
$h_1(y_1)$ & $\sqrt{2\pi}\frac{\sigma}{\sqrt{1-\alpha}}\exp(y_1^2/(2\frac{\sigma^2}{1-\alpha}))$ & $((1-\alpha)^{(1-\alpha)y_1+1}\gamma^{(1-\alpha)y_1}((1-\alpha)y_1)!)^{-1}$ & $\frac{\ell^{(1-\alpha)\ell}((1-\alpha)y_1)^{(1-\alpha)\ell-1}}{(1-\alpha)\Gamma((1-\alpha)\ell)}$  & $ \frac{1}{1-\alpha}{(1-\alpha)\ell \choose (1-\alpha)y_1}$ \\ \hline
$h_2(y_2)$ & $\sqrt{2\pi} \frac{\sigma}{\sqrt{\alpha}}\exp(y_1^2/(2\frac{\sigma^2}{\alpha}))$ & $(\alpha^{\alpha y_2+1}\gamma^{\alpha y_2}(\alpha y_2)!)^{-1}$ &  $\frac{\ell^{\alpha\ell}(\alpha y_2)^{\alpha\ell-1}}{\alpha\Gamma(\alpha\ell)}$ & $ \frac{1}{\alpha}{\alpha\ell \choose \alpha y_2}$ \\ \bottomrule
\end{tabular} 
\caption{\label{table:distributions} Examples of one-dimensional natural exponential family distributions $p(y|x)$ and their respective decompositions. These can be extended to higher dimensions by considering separable distributions $p(\y|\x)=\prod_{i=1}^{n} p(y_i|x_i)$, by $\eta(\x) = \sum_{i=1}^{n}\eta(x_i)$, $\phi(\x) = \sum_{i=1}^{n}\phi(x_i)$, $h(\y)=\prod_{i=1}^{n} h(y_i)$, $h_1(\y_1)=\prod_{i=1}^{n} h_1(y_{1,i})$ and $h_2(\y_2)=\prod_{i=1}^{n} h_2(y_{2,i})$.}
\end{table}

\newpage

\subsection*{Equivalence with SURE as $\alpha\to 0$}

\begin{equation}
\lim_{\alpha\to 0} \loss{GR2R-MSE}^{\alpha}(\y;f) = 
  \| f(\y) - \y\|^2_2  +  2 \sum_{i=1}^n \sum_{k\geq 1} (-1)^{k+1} a_k(y_i) \frac{1}{k!} \frac{\partial^k f_i}{\partial y_i^k}(\y) + \const. \label{eqn:gr2r_sure}
\end{equation}
where 
\begin{equation}
     a_k(y_i) = \lim_{\alpha\to 0} \mathbb{E}_{y_{2,i}|y_i,\alpha}(y_{2,i}-y_i)(\alpha y_{2,i})^{k}.
\end{equation}

\paragraph{Gaussian case.}
Based on the proposed re-corruption procedure for the Gaussian case, we have that the re-corruption of $\y_2$ in terms of $\y$ and the extra noise $\boldsymbol{\omega}$ as 
\begin{equation}
    \y_2 = \y - \sqrt{\frac{1-\alpha}{\alpha} } \boldsymbol{\omega}
\end{equation}
Analyze for $k=1$ 
\begin{equation}
    a_1(y_i) = \lim_{\alpha\to 0} \mathbb{E}_{y_{2,i}|y_i, \alpha}\{(y_{2,i}-y_i)(\alpha y_{2,i})^{1}\}
\end{equation}
for one element $y_2, y$
\begin{equation}
\begin{aligned}
        \lim_{\alpha\to 0} \mathbb{E}_{y_2|y, \alpha}\{(y_2-y)(\alpha y_2)\} &= \lim_{\alpha\to 0} \mathbb{E}_{\omega|y, \alpha}\{ \alpha( y - \sqrt{\frac{1 - \alpha}{\alpha}} \omega - y) (y - \sqrt{\frac{1 - \alpha}{\alpha}} \omega) \} \\
        &= \lim_{\alpha\to 0} \mathbb{E}_{\omega|y, \alpha}\{ - \alpha \sqrt{\frac{1 - \alpha}{\alpha}} \omega  ( y - \sqrt{\frac{1 - \alpha}{\alpha}} \omega ) \} \\ &= \lim_{\alpha\to 0} \mathbb{E}_{\omega|y, \alpha}\{ - \alpha \sqrt{\frac{1 - \alpha}{\alpha}} \omega y   + \alpha \frac{1 - \alpha}{\alpha} \omega^2 )   \}
        \\ &=  \lim_{\alpha\to 0} \mathbb{E}_{\omega|y, \alpha}\{ 
            - \sqrt{\alpha(1-\alpha)}\omega y + (1-\alpha)\omega^2
        \}
        \\ &= \lim_{\alpha\to 0}  (1 - \alpha) \sigma^2 = \sigma^2
\end{aligned}
\end{equation}
analyzing for $k>1$ we have that $a_{k}(y) \rightarrow 0$ since the $\alpha^{k-1}$ term dominates in the expression
\begin{equation}
   a_{k}(y) =  \lim_{\alpha\to 0} \mathbb{E}_{\omega|y, \alpha} 
    \Big\{ 
       ( - \sqrt{\alpha(1-\alpha)}\omega y + (1-\alpha)\omega^2 )  ( y -  \sqrt{\frac{1 - \alpha}{\alpha} } \omega )^{k-1} \alpha^{k-1} 
    \Big\}
\end{equation}
finally, substituting $a_{k}(y_i)$ in \eqref{eqn:gr2r_sure} for the Gaussian case we have that
\begin{equation}
    \lim_{\alpha\to 0} \loss{GR2R-MSE}^{\alpha}(\y;f) = 
     \| f(\y) - \y\|^2_2  +  2 \sigma^2 \sum_{i=1}^n \frac{\partial^k f_i}{\partial y_i^k}(\y) + \mathrm{const.}
\end{equation}

\paragraph{Poisson case.} Starting from $\mathcal{L}_{\text{GR2R-MSE}}^\alpha$ with $\y_1$ constructed in terms of $\y, \y_2$ and $\alpha$ as $\y_1 = (\y - \y_2 \alpha ) / (1 - \alpha)$ we have that

\begin{equation}
   \loss{GR2R-MSE}^{\alpha}(\y;f) = \mathbb{E}_{\y_2|\y}  \, \| f(\frac{\y-\y_2\alpha}{1-\alpha}) - \y\|^2_2  -\mathbb{E}_{\y_2|\y} \, 2 \sum_{i=1}^n \left(y_{2,i}-y_i\right) f_i(\frac{\y-\y_2\alpha}{1-\alpha}) + \mathrm{const,}
\end{equation}
evaluating $\lim_{\alpha\to 0} \loss{GR2R-MSE}^{\alpha}(\y;f) $

\begin{equation}
\begin{aligned}
    \lim_{\alpha\to 0} \loss{GR2R-MSE}^{\alpha}(\y;f) 
    &\propto \lim_{\alpha\to 0} \mathbb{E}_{\y_2|\y}  \, \| f(\frac{\y-\y_2\alpha}{1-\alpha}) - \y\|^2_2  -  \lim_{\alpha\to 0}\mathbb{E}_{\y_2|\y} \, 2 \sum_{i=1}^n \left(y_{2,i}-y_i\right) f_i(\frac{\y-\y_2\alpha}{1-\alpha}) \\
     &\propto \| f(\y) - \y\|^2_2  
    + 2 \lim_{\alpha\to 0}  \mathbb{E}_{\y_2|\y} \, \Big( \sum_{i=1}^n y_i  f_i(\frac{\y-\y_2\alpha}{1-\alpha}) -  y_{2,i}  f_i(\frac{\y-\y_2\alpha}{1-\alpha})  \Big)\\
    &\propto  \| f(\y) - \y\|^2_2 + 2 \sum_{i=1}^n \Big( y_i f_i(\y) -   \lim_{\alpha\to 0} \mathbb{E}_{\y_2|\y}  y_{2,i}  f_i(\frac{\y-\y_2\alpha}{1-\alpha}) 
    \Big)
\end{aligned}
\end{equation}

Recall that $\y=\gamma \z $ and $\y_2 = \gamma \boldsymbol{\omega} /\alpha$ with $\boldsymbol{\omega} \sim \text{Bin}(\z,\alpha)$. Defining the function $g_{i,\alpha}: \omega_{i}\mapsto f_i(\frac{\y-\gamma\boldsymbol{\omega}}{1-\alpha})$, we have that the second term is 
\begin{align}
    \lim_{\alpha\to 0} \mathbb{E}_{\y_2|\y}  
 y_{2,i}  f_i(\frac{\y-\y_2\alpha}{1-\alpha}) &=   \lim_{\alpha\to 0} \mathbb{E}_{\boldsymbol{\omega}_{-i}|y_i} \mathbb{E}_{\omega_i|y_i}  \gamma \frac{\omega_i}{\alpha}   f_i(\frac{\y-\gamma
 \boldsymbol{\omega}}{1-\alpha})  \\ &= \lim_{\alpha\to 0} \sum_{k=1}^{z_i}  \gamma  {z_{i} \choose k } \alpha^{k-1} (1-\alpha)^{z_i-k} k\, \mathbb{E}_{\boldsymbol{\omega}_{-i}|y_i}  g_{i,\alpha}(k) \\
 &= \lim_{\alpha\to 0} \Big(\gamma  z_i \,  (1-\alpha)^{z_i-1} \mathbb{E}_{\boldsymbol{\omega}_{-i}|z_i} g_{i,\alpha}(1)    + \sum_{k=2}^{z_i} \gamma  { z_{i}  \choose k } \alpha^{k-1}(1-\alpha)^{z_i-k} k \, \mathbb{E}_{\boldsymbol{\omega}_{-i}|y_i}  g_{i,\alpha}(k)  \\
 &\propto \lim_{\alpha\to 0} \Big( \gamma z_i \,  (1-\alpha)^{z_i-1} \mathbb{E}_{\boldsymbol{\omega}_{-i}|y_i}  g_{i,\alpha}(1)    + \mathcal{O}(\alpha) \Big)   \\ &\propto  \gamma z_i \lim_{\alpha\to 0} \mathbb{E}_{\boldsymbol{\omega}_{-i}|y_i} g_{i,0}(1)  \\ &\propto y_i f_i(\y-\gamma\boldsymbol{e}_i),
\end{align}
where $\boldsymbol{e}_i \in \mathbb{R}^n$ is the vector with $i$-th entry in 1 and with all others in 0. Thus, plugging in this result, we have 

\begin{equation}
     \lim_{\alpha\to 0} \loss{GR2R-MSE}^{\alpha}(\y;f)  = \| f(\y) - \y\|^2_2  + 2 \sum_{i=1}^n y_i \Big( f_i(\y) -  f_i( \y - \gamma\boldsymbol{e}_i ) \Big) + \mathrm{const.}
\end{equation}

\paragraph{Gamma case.} Based on the proposed re-corruption procedure for the Gamma case, we have that the re-corruption of $\y_2$ in terms of $\y$ and the extra noise $\boldsymbol{\omega} \sim  \text{Beta} (\ell \alpha, \ell (1-\alpha))$ as \begin{equation}
    \y_2 = \frac{\boldsymbol{\omega}}{\alpha} \y
\end{equation}
then, replacing in the expression of $a_k(y_i)$ for one element $y_2, y$ \begin{equation}
\begin{aligned}
     a_k(y) &= \lim_{\alpha\to 0} \mathbb{E}_{y_{2}|y, \alpha}\{(y_{2}-y)(\alpha y_{2})^{k}\} = \lim_{\alpha\to 0}   \mathbb{E}_{\omega |y, \alpha} \{ ( \frac{\omega}{\alpha} y - y ) (\omega y ) ^k \} \\
     &= \lim_{\alpha\to 0}   \mathbb{E}_{\omega|\alpha} \{
     \frac{\omega^{k+1}}{\alpha} y^{k+1} -\omega^{k} y^{k+1} \} \\
      &=  y^{k+1}  \lim_{\alpha\to 0}  \Big( \frac{1}{\alpha}  \mathbb{E}_{\omega|\alpha} \{
     \omega^{k+1} \} - \mathbb{E}_{\omega|\alpha}\{\omega^{k}\} \Big) \\
    \end{aligned}
\end{equation} 
The $k$th moment of $\omega$ can be expressed recursively as 
\begin{equation}
    \mathbb{E}\{ \omega^{k+1} \} = \frac{\ell\alpha+k-1}{\ell+k-1} \mathbb{E}\{\omega^k \} 
\end{equation}
then 
\begin{align}
 \lim_{\alpha\to 0}  \Big( \frac{1}{\alpha}  \mathbb{E}_{\omega|\alpha} \{
     \omega^{k+1} \} - \mathbb{E}_{\omega|\alpha}\{\omega^{k}\} \Big) &= \lim_{\alpha\to 0} \mathbb{E}_{\omega|\alpha}\{\omega^{k}\} \Big( \frac{1}{\alpha}  \frac{\ell\alpha+k-1}{\ell+k-1} - 1 \Big) \\
     &= \lim_{\alpha\to 0} \mathbb{E}_{\omega|\alpha}\{\omega^{k}\}   \frac{\ell\alpha+k-1 - \alpha(\ell+k-1) }{\alpha(\ell+k-1)}  \\
     &= \lim_{\alpha\to 0} \mathbb{E}_{\omega|\alpha}\{\omega^{k}\}  \frac{(k-1)(1-\alpha) }{\alpha(\ell+k-1)} \\
     &= \lim_{\alpha\to 0} (\prod_{r=0}^{k-1} \frac{\alpha \ell +r }{\ell+r} ) \frac{(k-1)(1-\alpha) }{\alpha(\ell+k-1)} \\
     &= \lim_{\alpha\to 0} (\prod_{r=1}^{k-1} \frac{\alpha \ell +r }{\ell+r} )  \frac{\alpha \ell}{\ell}\frac{(k-1)(1-\alpha) }{\alpha(\ell+k-1)} \\
       &= \lim_{\alpha\to 0} (\prod_{r=1}^{k-1} \frac{\alpha \ell +r }{\ell+r} ) \frac{(k-1)(1-\alpha) }{(\ell+k-1)} \\
       &= (\prod_{r=1}^{k-1} \frac{r }{\ell+r} ) \frac{(k-1) }{(\ell+k-1)}\\       
       &= \frac{(k-1)!\, \Gamma(\ell)}{\Gamma(\ell+k)} \frac{\ell(k-1) }{(\ell+k-1)} \\
\end{align}

Finally, substituting $a_{k}(y_i)$ in \eqref{eqn:gr2r_sure} for the Gamma case we have that
\begin{equation}
     \lim_{\alpha\to 0} \loss{GR2R-MSE}^{\alpha}(\y;f) = 
     \| f(\y) - \y\|^2_2  +  2 \sum_{i=1}^n \sum_{k\geq1} 
     \frac{\ell(k-1)}{k(\ell+k-1)}
     \frac{(-y_i)^{k+1}\Gamma(\ell)}{\Gamma(\ell+k)} \frac{\partial^k f_i}{\partial y_i^k}(\y) + \mathrm{const.}
\end{equation}

\section{Experimental details} \label{app: experiments}
The maximum-entropy sampling strategy, detailed below, is employed to generate noise that ensures the third moment is preserved in the experiment described in Section 4.1. Non-Gaussian Additive Noise, in the main paper.

\subsection{Maximum-entropy sampling}
Consider a random variable $z$ with $\mu_i = \mathbb{E} \, z^{i}$ the desired moments of order $i=1,\dots,k$.
We obtain maximum entropy samples verifying the desired moments up to order $k$ by minimizing~\cite{bruna2013audio}
\begin{equation}
    \arg \min_{\z}  \sum_{i=0}^{k} \|  \frac{1}{n}\sum_{j=1} z_j^{i} - \mu_{i}\|^2_2
\end{equation}
via gradient descent where we initialize $\boldsymbol{z}\sim \mathcal{N}(\mu_{1}\boldsymbol{1}, \boldsymbol{I} (\mu_2-\mu_{1}^2))$. The optimization is stopped when the relative error is small, i.e.,
$$ \frac{\frac{1}{n}|\sum_{j=1} z_j^{i} - \mu_{i}|}{|\mu_i| } < 0.1$$
for all $i=1,\dots,k$.

\newpage
\section{Additional Simulations and Results}

\subsection{Effect of the re-corruption hyper-parameter $\alpha$.}

We evaluate the performance of the proposed GR2R loss on the PSNR metric when examining the effect of the re-corruption parameter $\alpha$ on three noise distributions: Poisson, Gamma, and Gaussian. Specifically, the experimental setup consists of training the DnCNN model architecture by minimizing the proposed loss $\mathcal{L}_{\text{GR2R-MSE}}^\alpha$ for different values of $\alpha$ on the DIV2K dataset. All experiments share the same training configuration:  Adam optimizer, with an initial learning rate of 1e-4 and 250 training epochs. For the noise model parameters, we set $\gamma=0.5$ for the Poisson experiment, $\ell=5$ for the Gamma experiment, and $\sigma=0.1$ for the Gaussian experiment. 

\begin{figure}[!h]
    \centering 
    \includegraphics[width=\linewidth]{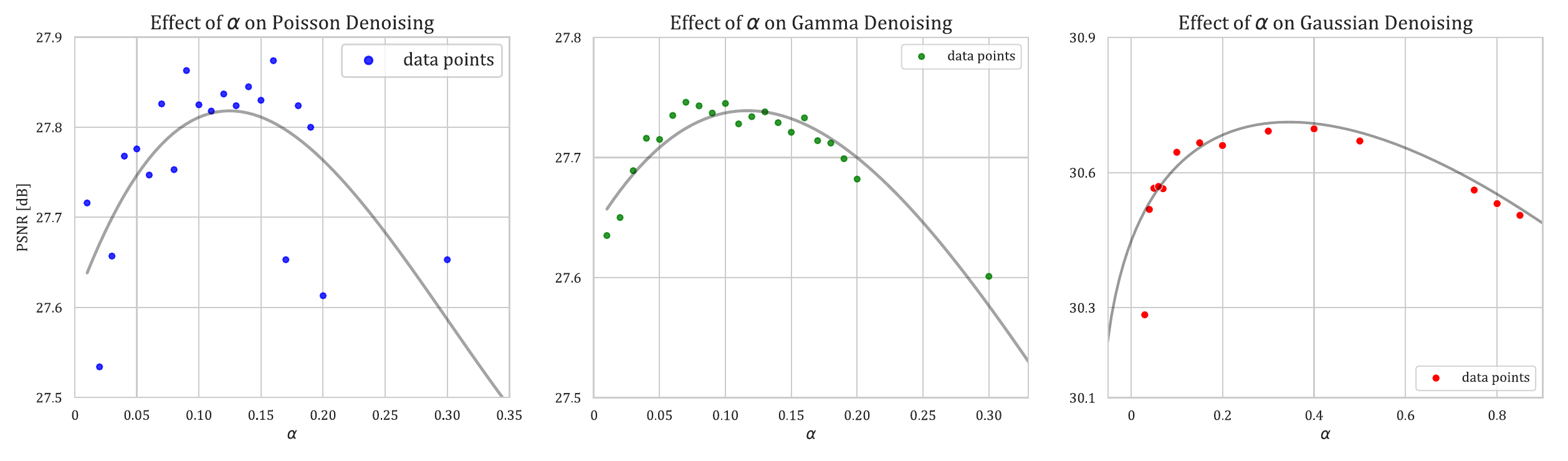} \vspace{-2em}
    \caption{Effect of $\alpha$ parameter for different noise distributions. he results indicate that the optimal $\alpha$ parameter consistently lies within the range of $0.1$ to $0.3$ across all tested scenarios.}
    \label{fig:alphatest}
\end{figure}

We test the GR2R loss for $\alpha$ values in the interval [0.1, 3.5] for Poisson and Gamma and in the interval [0.1, 0.9] for Gaussian. A scatter plot is shown in Figure~\ref{fig:alphatest} for all noise distributions tested, with the trends highlighted by polynomial fitting. A trade-off between the value of $\alpha$ and the PSNR score can be observed, where low values of $\alpha$ indicated less SNR in $\y_1$ and higher SNR in $\y_2$. For the Poisson and Gamma distributions, the optimal values of the re-corruption parameter $\alpha$ appear to be approximately $\alpha=0.12$, while for the Gaussian distribution, the preferred value seems to be $\alpha=0.3$. Furthermore, although the performance of the GR2R loss is sensitive to the choice of the re-corruption parameter $\alpha$. The disparity between the highest and lowest PSNR scores is less than 0.2 dB for the Gamma and Gaussian distributions and less than 0.6 dB for Gaussian noise.

\vspace{-2em}

\subsection{Log-Rayleigh Noise}

\begin{wrapfigure}{R}{0.42\textwidth} \vspace{-2em}
    \includegraphics[width=0.42\textwidth]{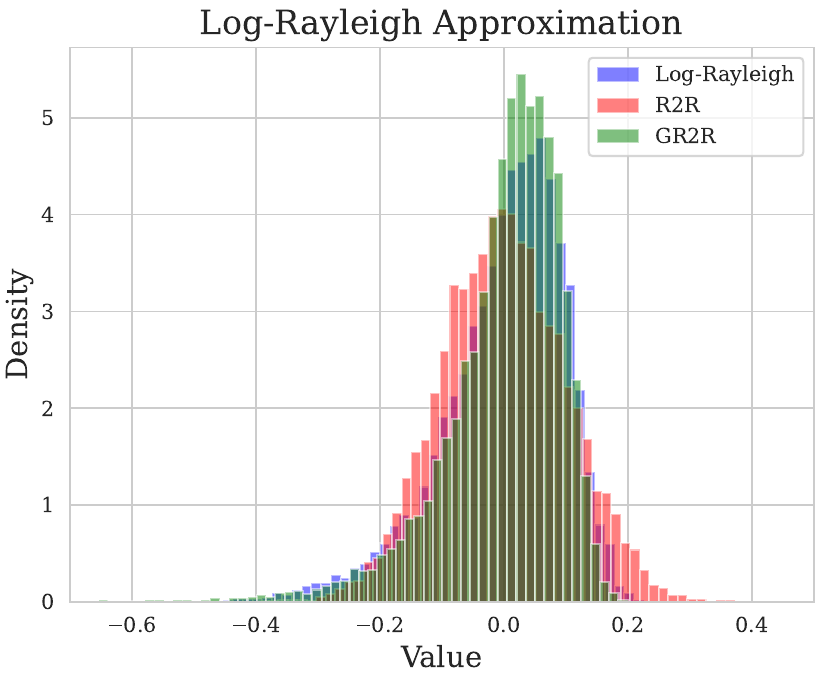}
    \caption{Histogram of noise estimations.}
    \label{fig:histogram}
\end{wrapfigure}

\begin{figure}[!h]
    \centering
    \includegraphics[width=\linewidth]{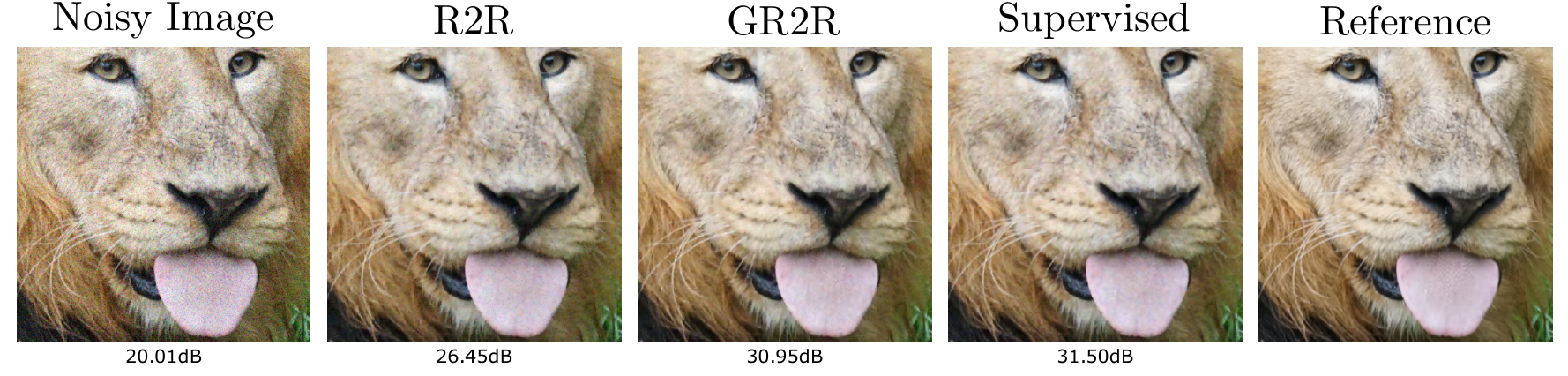}
    \includegraphics[width=\linewidth]{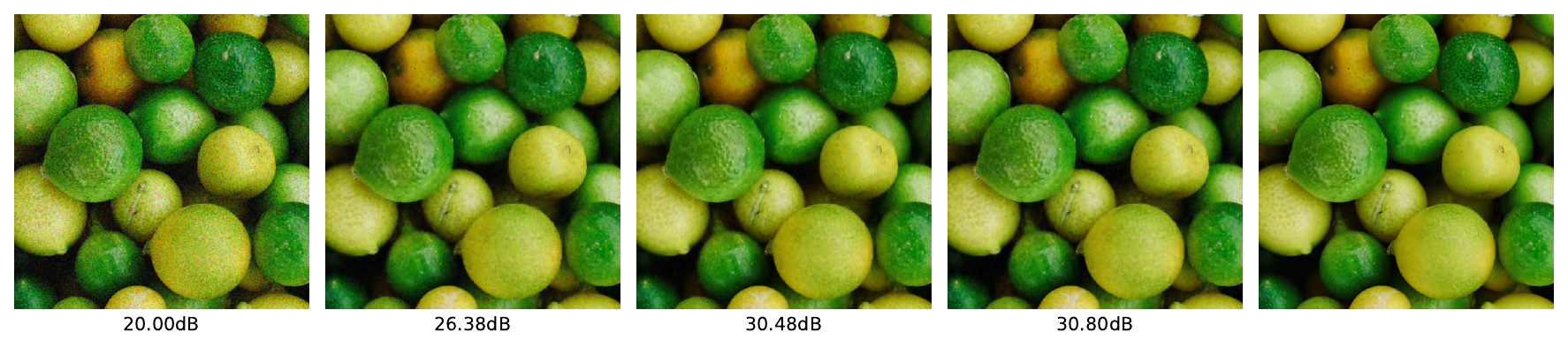}
    \includegraphics[width=\linewidth]{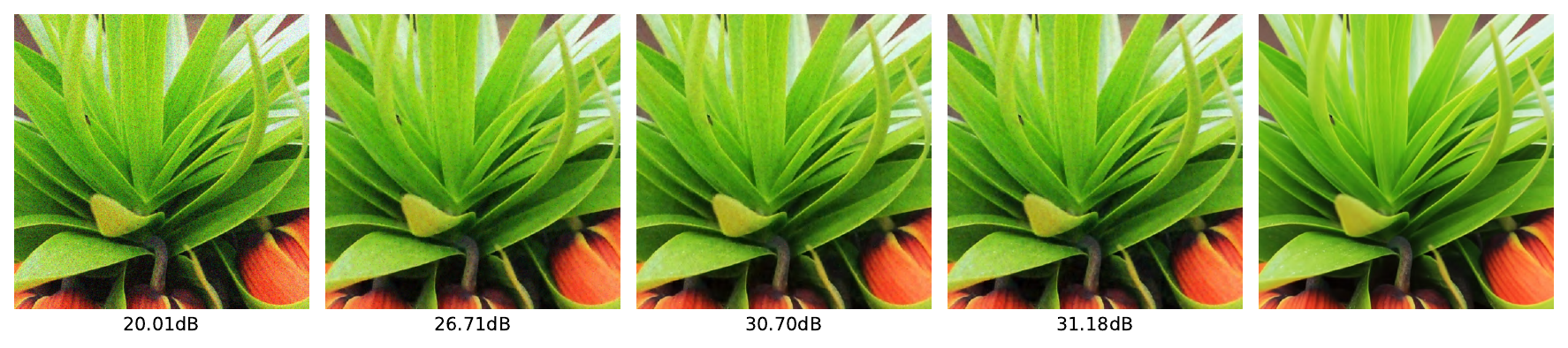}
    \includegraphics[width=\linewidth]{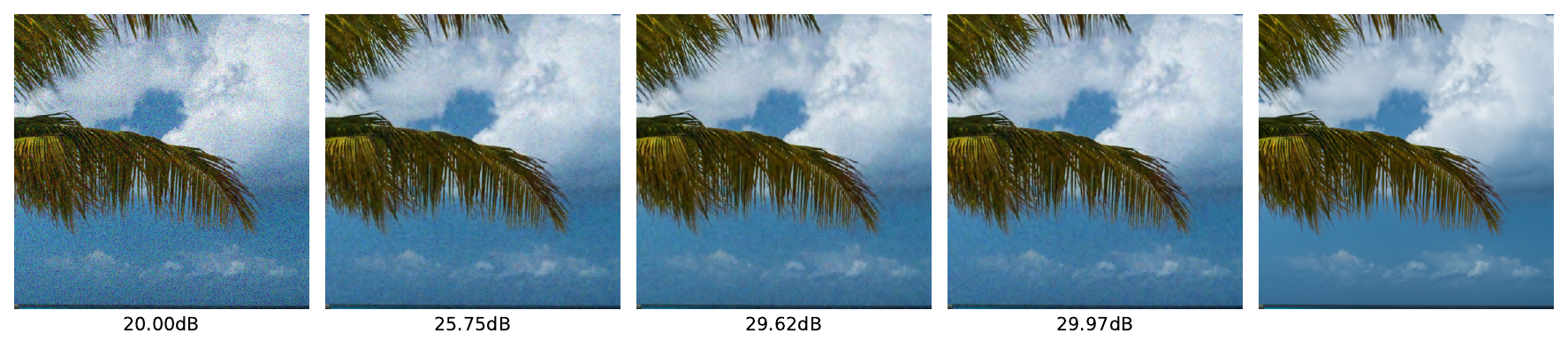}
    \includegraphics[width=\linewidth]{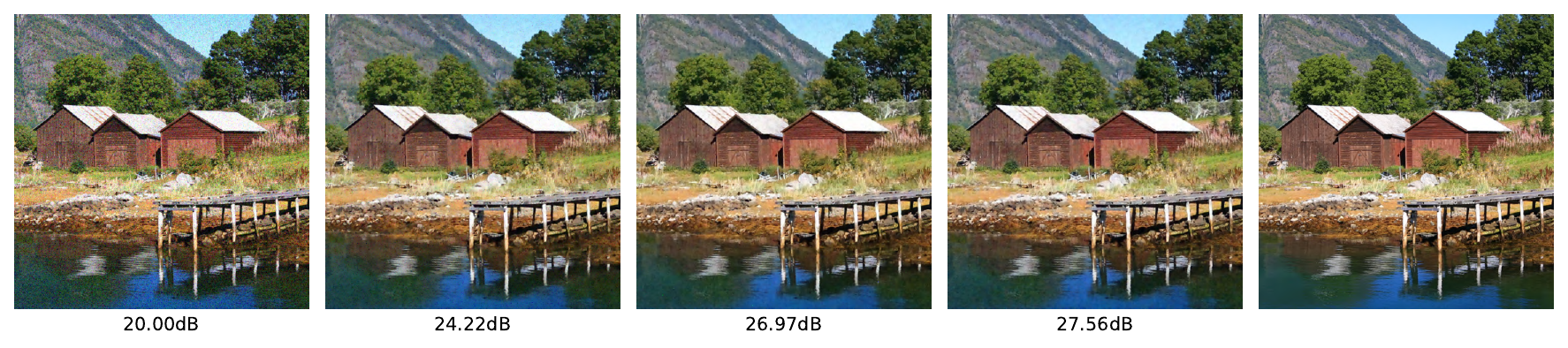}
    \caption{Visual Results for a Log-Rayleigh Noise with a standard deviation of $\sigma=0.1$.}
    \label{fig:lograyleigh}
\end{figure}

In addition to the numerical comparisons presented in the main manuscript between R2R (matching second-order moment) and the proposed GR2R (matching third-order moment), presented in Section 4.1 in the main document, this section offers further elaboration on the experimental setups, as well as visual analyses of the noise estimation compared to the restored images. The training configuration consists of the DnCNN model along 100 epochs with a batch size of 15 with an initial learning rate of 5e-4 with the Adam optimizer in the DIV2K dataset. Figure~\ref{fig:histogram} displays a histogram comparing the original Log-Rayleigh noise, which was utilized to corrupt the images, with the estimated additional noise provided by R2R and GR2R. It can be observed that extending the moment matching to the third moment significantly enhances the accuracy of the noise distribution estimation compared to matching only until the second moment. Restored images are presented in Figure~\ref{fig:lograyleigh}, which demonstrate \\the effect of matching the third moment in image denoising.

\newpage

\newpage

\subsection{Additional Results}
The following subsections present results of the PSNR mean and standard deviation obtained for the different methods for Poisson, Gamma, and Gaussian distributions. Each subsection also shows additional visual results. \vspace{-1em}

\subsubsection{Poission Noise}
\vspace{-2em}
\begin{table*}[!h]
    \centering \small
    \caption{PSNR results on Poisson noise. GR2R-NLL stands for the proposed GR2R with Negative Log-Likelihood.} 
    
    \begin{tabular}{l|| cccc|c } \hline
    Poisson Noise & \multicolumn{4}{c}{Methods}                                                             \\ \hline  
    Noise Level ($\gamma$)  & PURE~\cite{le2014unbiased}            & Neigh2Neigh~\cite{huang2021neighbor2neighbor}        & GR2R-NLL (ours)  & GR2R-MSE (ours)        & Supervised-MSE    \\  \hline \hline
    0.01          
    &  32.69$\pm$2.13  & 33.37$\pm$2.20  & 33.90$\pm$2.26 & \underline{33.92$\pm$2.20} & \textbf{33.96$\pm$2.23} \\ 
    0.1           &  24.37$\pm$1.89   & 28.27$\pm$2.60  & 28.30$\pm$2.65  & \underline{28.35$\pm$2.64} & \textbf{28.39$\pm$2.65} \\ 
    0.5           & 22.98$\pm$1.53    & 24.90$\pm$2.68  & \underline{25.07$\pm$2.71}  & 24.69$\pm$2.74 & \textbf{25.32$\pm$2.75}  \\ 
    1.0             & 17.94$\pm$1.13   & 23.56$\pm$2.67  & \underline{23.69$\pm$2.70} &  23.49$\pm$2.71 & \textbf{23.85$\pm$2.72}  \\ \bottomrule
    \end{tabular} 
    \label{tab:results_poisson} \vspace{-1em}
\end{table*}
\vspace{-1em}
\begin{figure}[!h]
    \centering
    \includegraphics[width=0.9\linewidth]{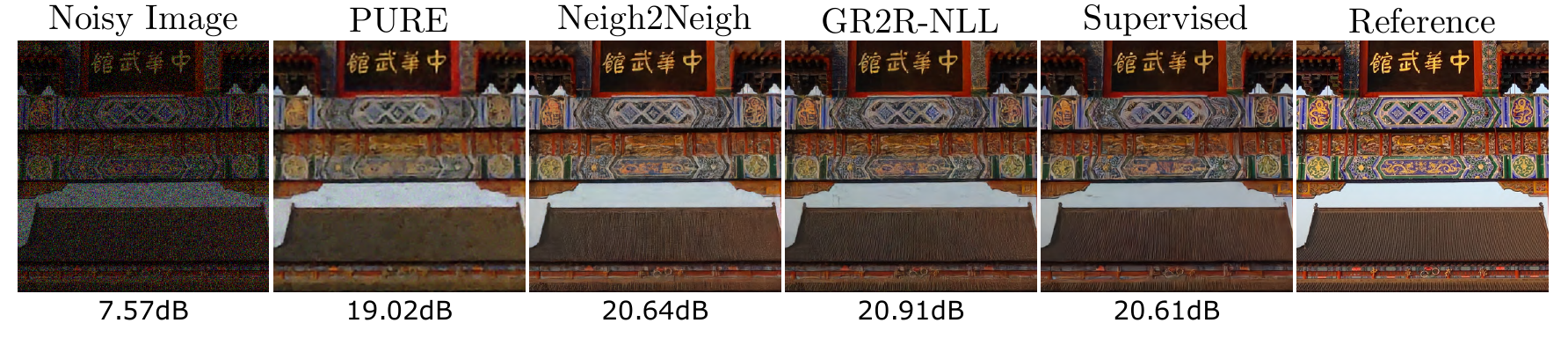} 
    \includegraphics[width=0.9\linewidth]{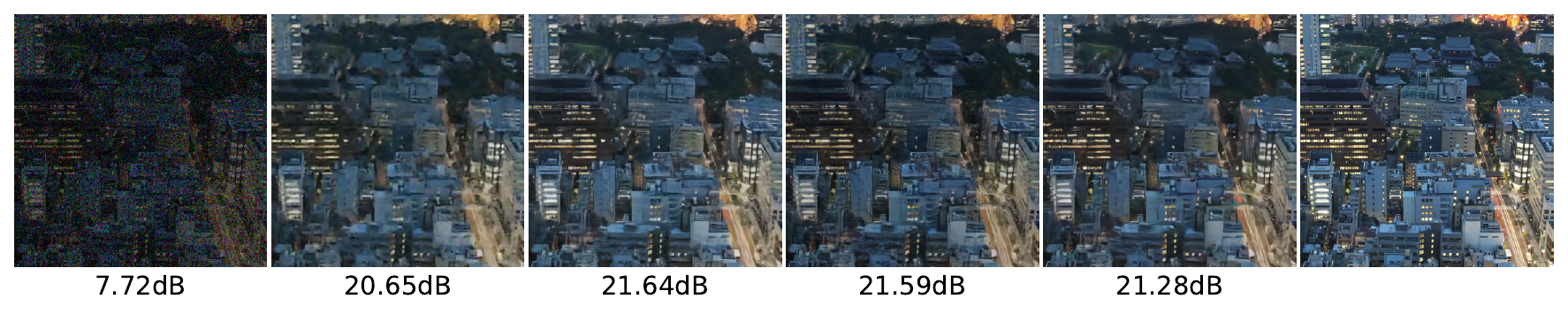}
    \includegraphics[width=0.9\linewidth]{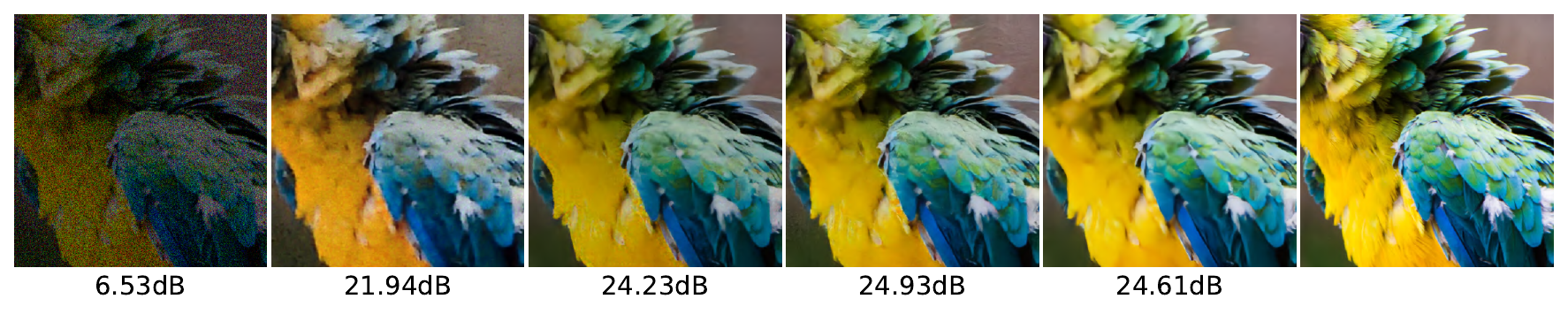}
    \includegraphics[width=0.9\linewidth]{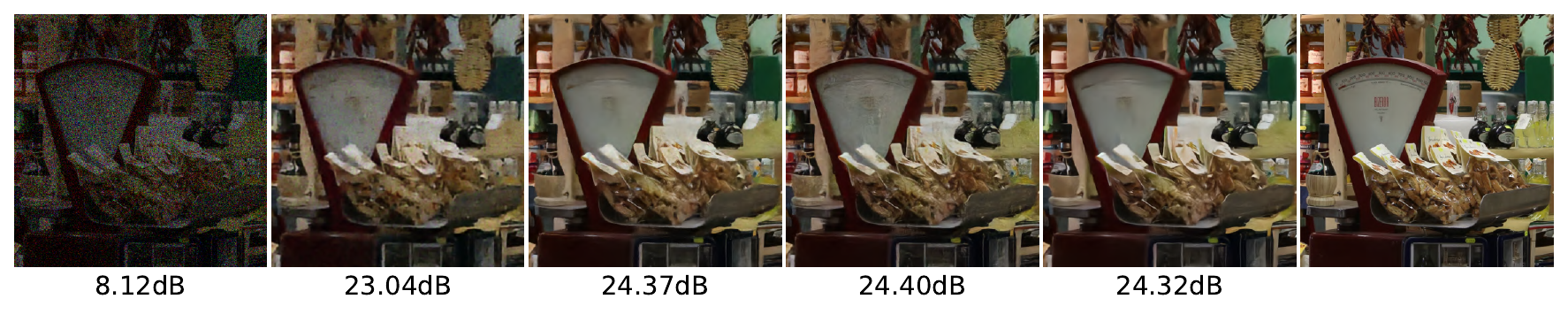}
    \includegraphics[width=0.9\linewidth]{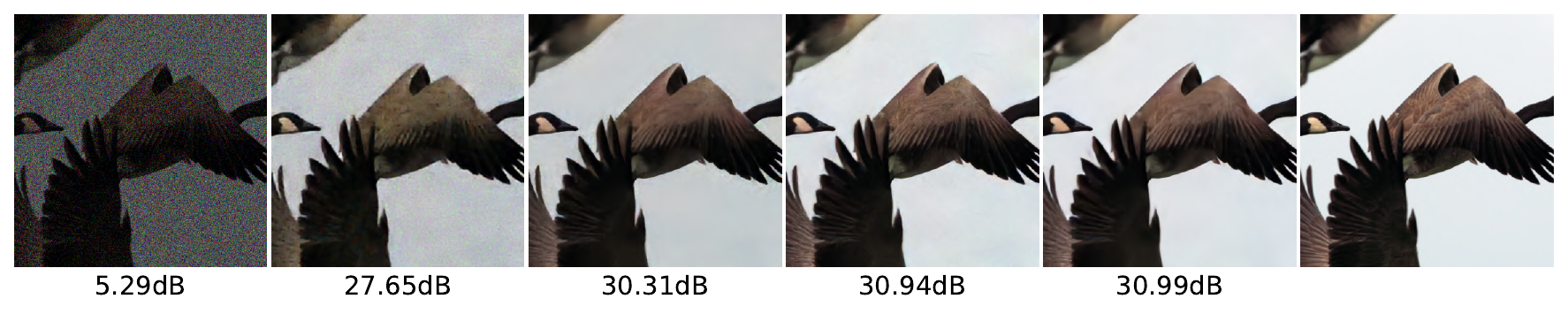} \vspace{-1.5em}
    \caption{Poisson Denoising in DIV2K Dataset.}
    \label{fig:enter-label}
\end{figure}


\subsubsection{Gamma Noise}

\begin{table}[!h]
    \centering 
    \caption{PSNR results on Gamma noise. GR2R-NLL stands for the proposed GR2R with Negative Log-Likelihood.} 
    \begin{tabular}{l|| ccc|c } \hline
    Gaussian Noise & \multicolumn{4}{c}{Methods}                                                             \\ \hline  
    Number of looks ($\ell$)   & Neigh2Neigh~\cite{huang2021neighbor2neighbor}            & GR2R-NLL (ours)       & GR2R-MSE (ours)         & Supervised-MSE    \\  \hline \hline
    30         &30.34$\pm$1.60   & 30.43$\pm$1.61  &  \underline{31.58$\pm$1.72}& \textbf{31.86$\pm$1.73} \\ 
    15         & 28.56$\pm$1.58  & 28.71$\pm$1.59  &  \underline{29.55$\pm$1.68}& \textbf{29.76$\pm$1.70}  \\ 
    5          & 25.71$\pm$1.53  & 25.79$\pm$1.49  &  \underline{26.35$\pm$1.57}& \textbf{26.72$\pm$1.62} \\ 
    1          & 22.19$\pm$1.40  & 22.19$\pm$1.34  &  \underline{22.38$\pm$1.40}& \textbf{22.56$\pm$1.44}  \\ \bottomrule
    \end{tabular} 
    \label{tab:results_gamma} \vspace{-1em}
\end{table}

\begin{figure}[!h]
    \centering
    \includegraphics[width=0.95\linewidth]{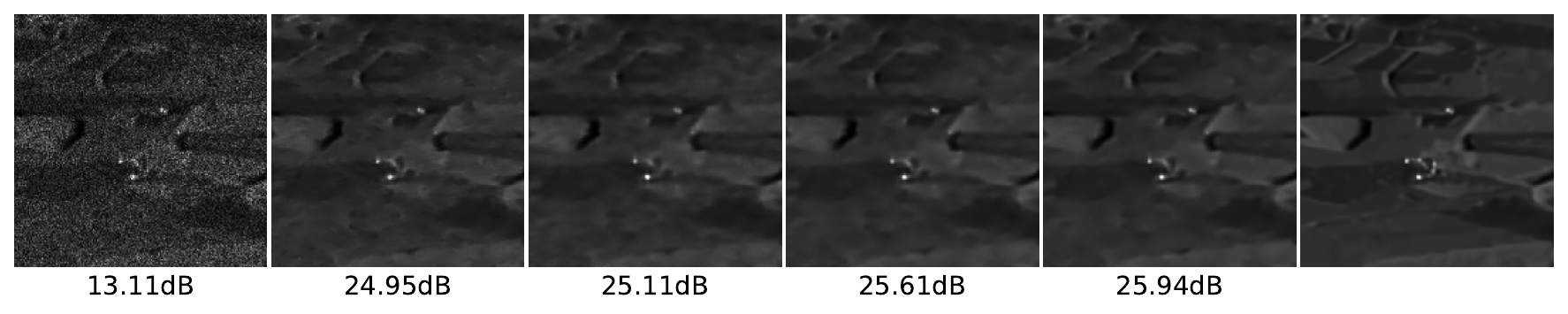}
    \includegraphics[width=0.95\linewidth]{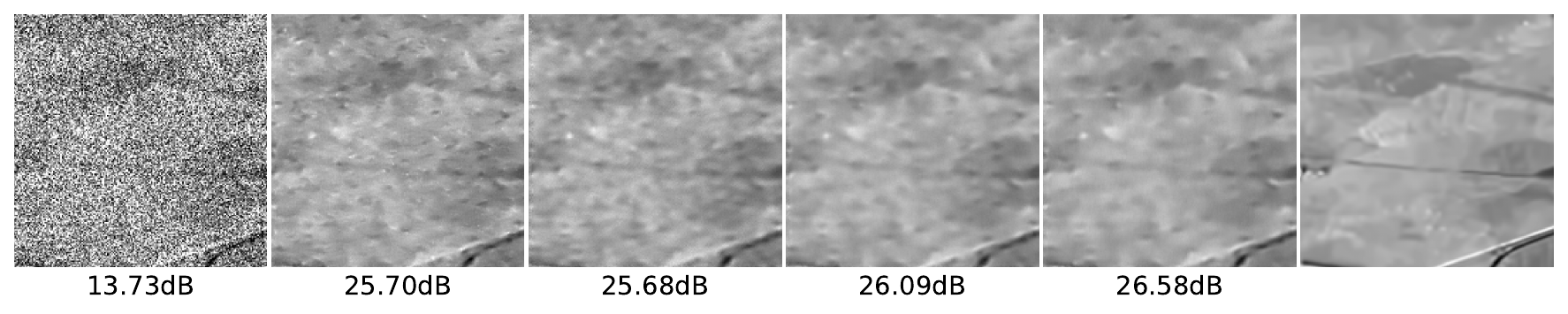}
    \includegraphics[width=0.95\linewidth]{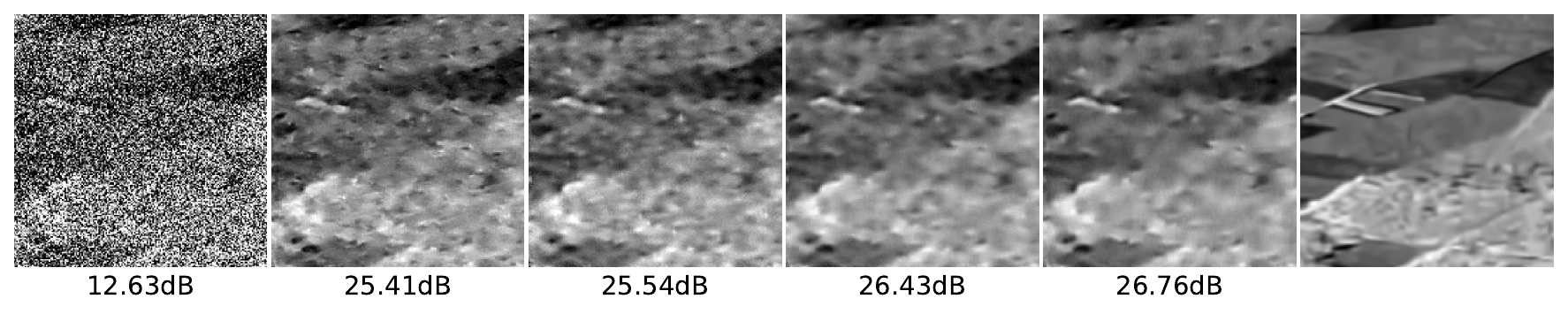}
    \includegraphics[width=0.95\linewidth]{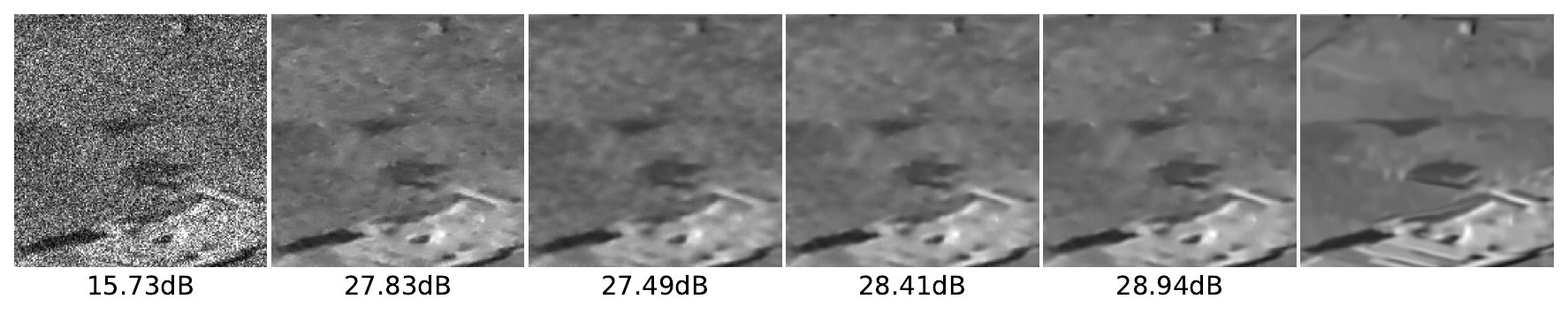}
    \includegraphics[width=0.95\linewidth]{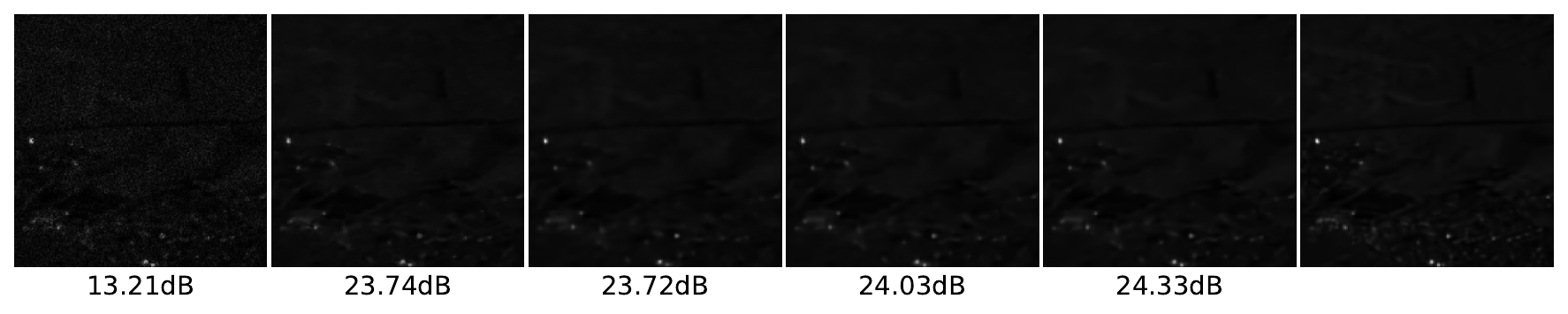} \vspace{-1em}
    \caption{Gamma Denoising in SARDataset.}
    \label{fig:enter-label}
\end{figure}


\newpage

\subsubsection{Gaussian Noise}

\begin{table*}[!h]
    \centering
    \caption{PSNR results for Gaussian noise. For this case, the MSE and NLL variants of GR2R are the same.}
    \begin{tabular}{l||cccc|c } \hline
    Gaussian noise & \multicolumn{4}{c}{Methods}                                                             \\ \hline  
    Noise Level ($\sigma$) & Noise2Score~\cite{kim2021noise2score}  & SURE~\cite{ramani2008monte}            & Neigh2Neigh~\cite{huang2021neighbor2neighbor}        & GR2R (ours)         & Supervised-MSE    \\  \hline \hline
    0.05         &  34.42$\pm$1.16 & 35.31$\pm$1.43  &  35.07$\pm$1.41  &  \underline{35.38$\pm$1.47} & \textbf{35.41$\pm$1.47} \\ 
    0.1          &  31.02$\pm$0.74 & 32.76$\pm$1.22   &  32.57$\pm$1.22  &  \underline{33.03$\pm$1.29} & \textbf{33.14$\pm$1.28}  \\ 
    0.2          &  29.34$\pm$0.62 & 29.77$\pm$1.02   &  29.73$\pm$1.05  &  \underline{30.24$\pm$1.05} & \textbf{30.38$\pm$1.05} \\ 
    0.5          &  22.94$\pm$0.65 & 25.52$\pm$1.02 & 25.61$\pm$0.99  & \underline{25.81$\pm$0.97} &  \textbf{25.93$\pm$0.94}  \\ \bottomrule
    \end{tabular}      
    \label{tab:results_gaussian} \vspace{-1em}
\end{table*}

\begin{figure}[!h]
    \centering
    \includegraphics[width=\linewidth]{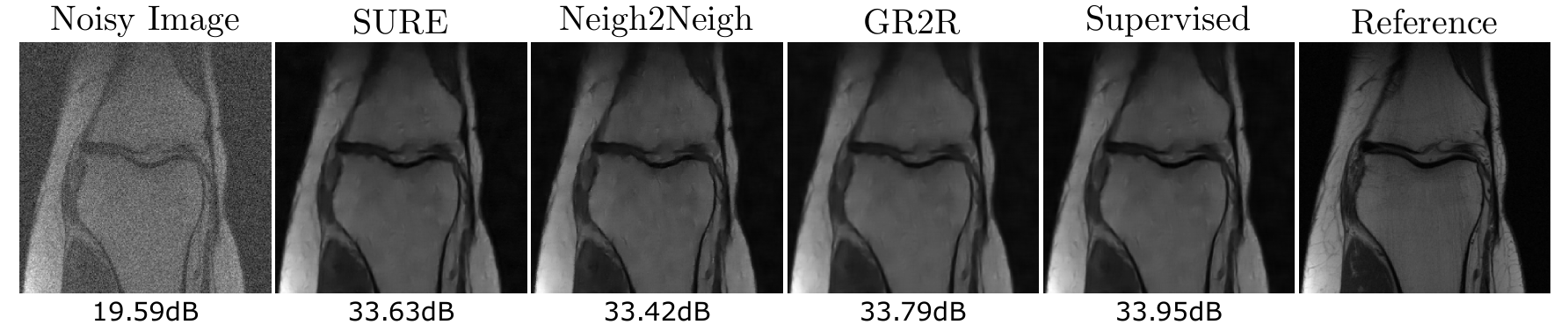}
    \includegraphics[width=\linewidth]{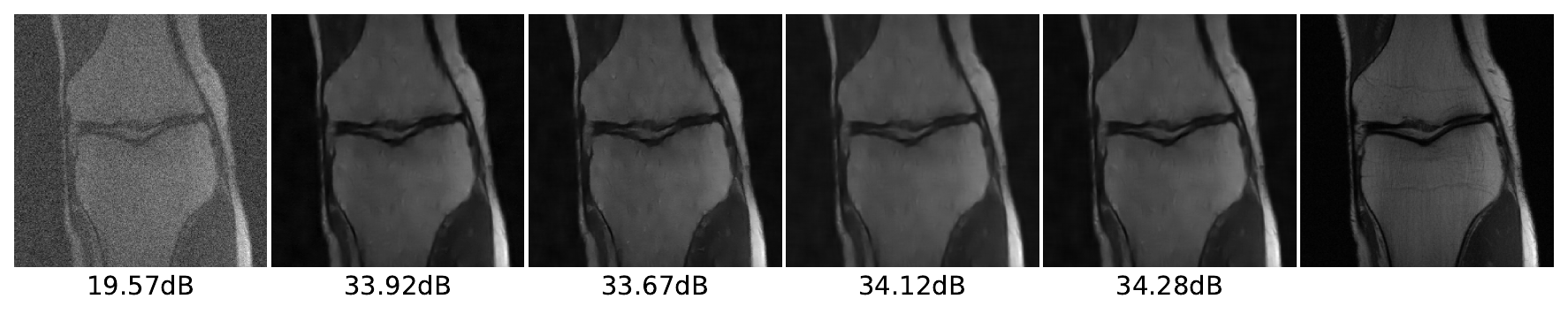}
    \includegraphics[width=\linewidth]{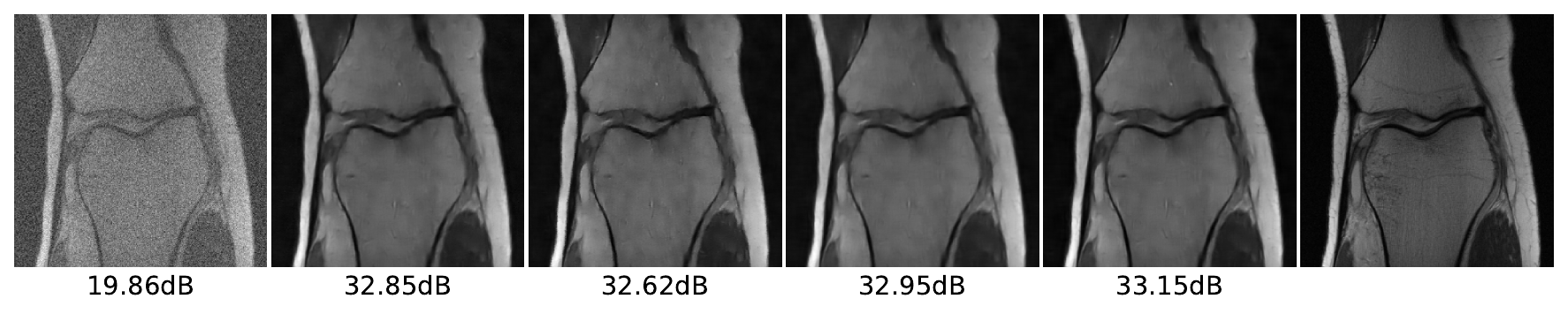}
    \includegraphics[width=\linewidth]{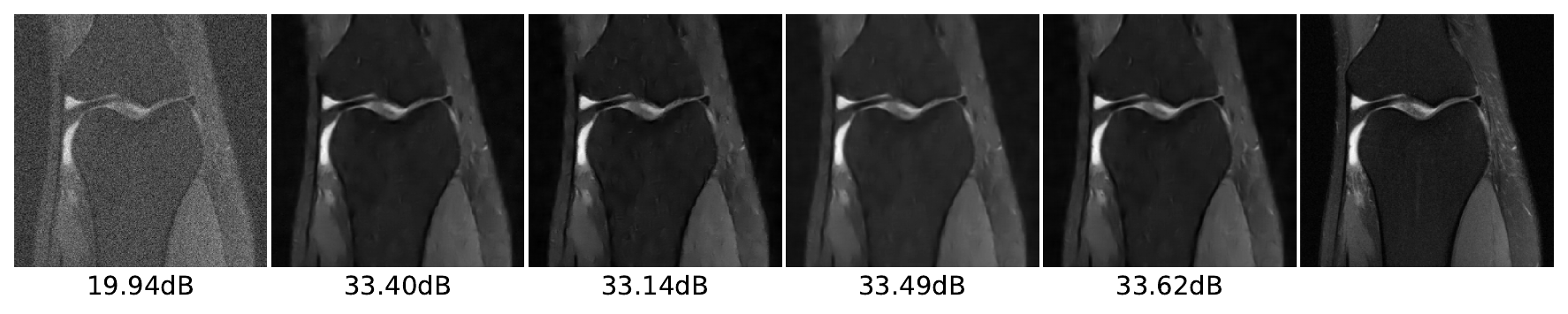}
    \includegraphics[width=\linewidth]{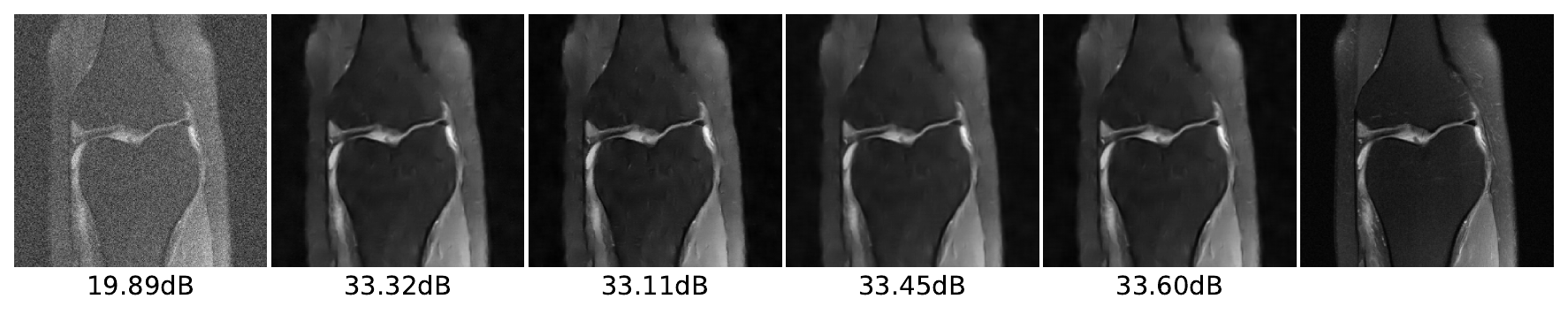} \vspace{-2em}
    \caption{Gaussian Denoising in MRI Dataset.}
    \label{fig:enter-label}
\end{figure}


\newpage

\section{General Inverse Problems}
This section extends the results of the self-supervised inpainting (Section 5 of the main paper ) for Poisson, Gamma, and Gaussian using DIV2K Dataset. \vspace{-1em}

\begin{table*}[!h]
    \centering
    \caption{PSNR/SSIM results for different noise models on inpainting in DIV2K dataset.}
    \begin{tabular}{l||ccc|c } \hline
     & \multicolumn{4}{c}{Methods}                                                             \\ \hline  
    Noise Model  & EI~\cite{chen2021equivariant}  & REI~\cite{chen2022robust}            & GR2R~(ours)        &  Supervised-MSE    \\  \hline \hline
    Poisson \;\;$\gamma=0.5$        &  22.53/0.627 & 27.05/0.777  &  \underline{27.41/0.791}  &   \textbf{28.42/0.832}  \\ 
    Gamma    \;\;$\ell=5$      &  17.06/0.467 & -   &  \underline{26.81/0.784}  &  \textbf{27.12/0.802} \\ 
    Gaussian $\sigma=0.1$    &  23.68/0.671 &  29.53/0.853   & \underline{29.58/0.854}  &  \textbf{29.93/0.866}   \\ \bottomrule
    \end{tabular}      
    \label{tab:results_gaussian} \vspace{-1em}
\end{table*}

\begin{figure}[!h]
    \centering
    \includegraphics[width=\linewidth]{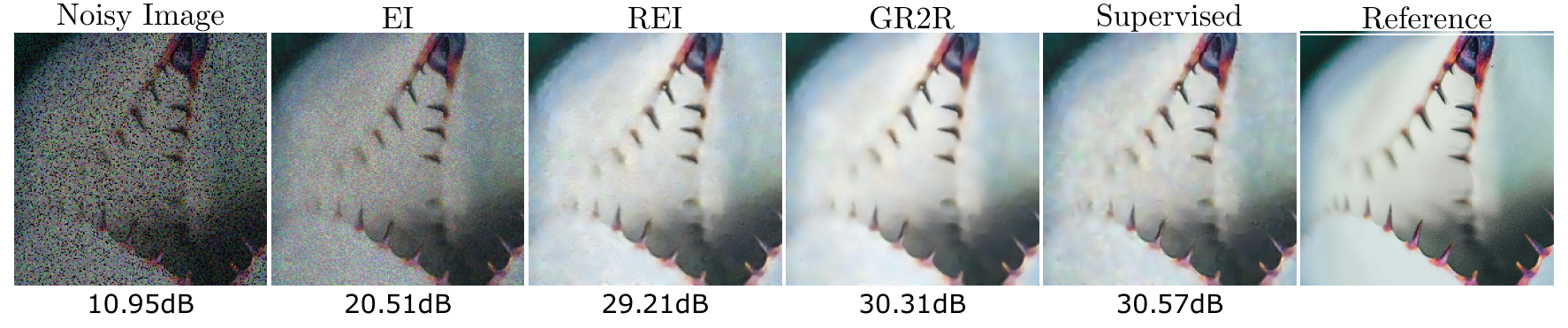}
    \includegraphics[width=\linewidth]{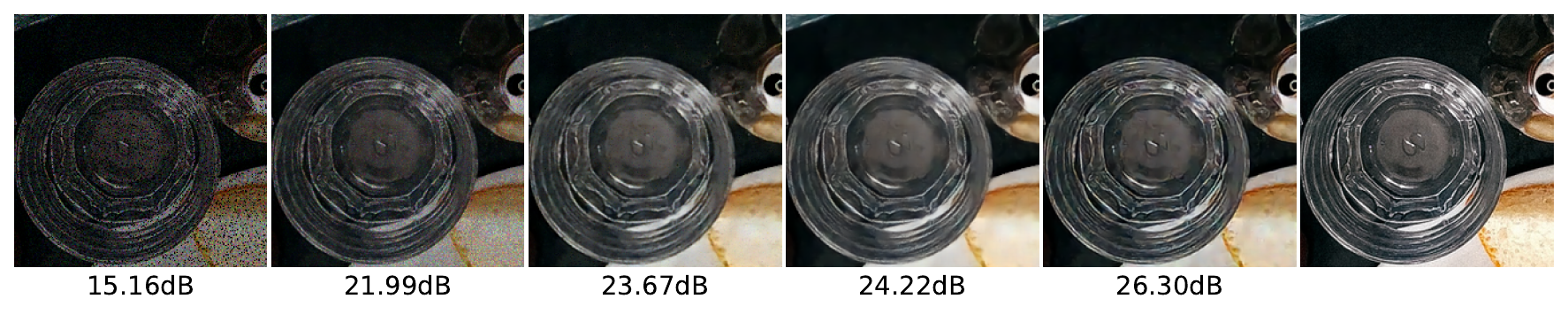}
    \includegraphics[width=\linewidth]{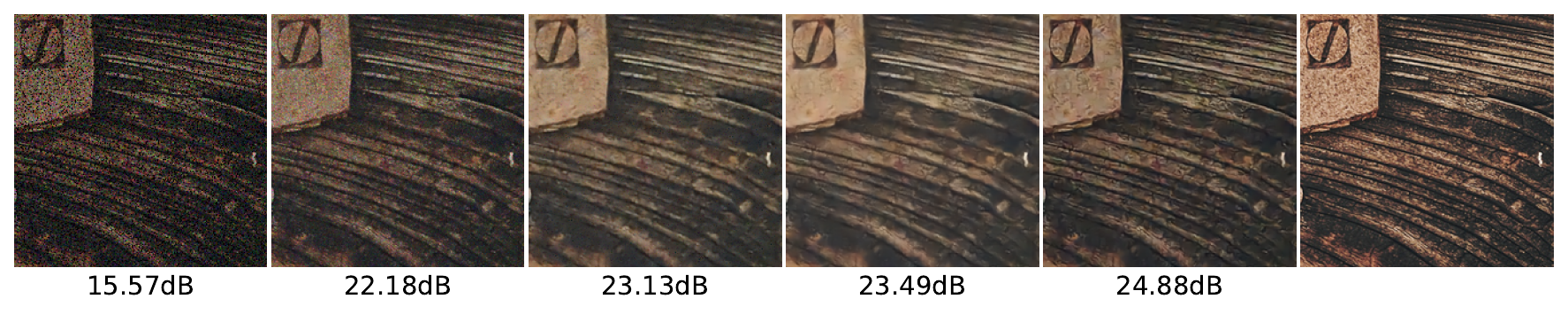}
    \includegraphics[width=\linewidth]{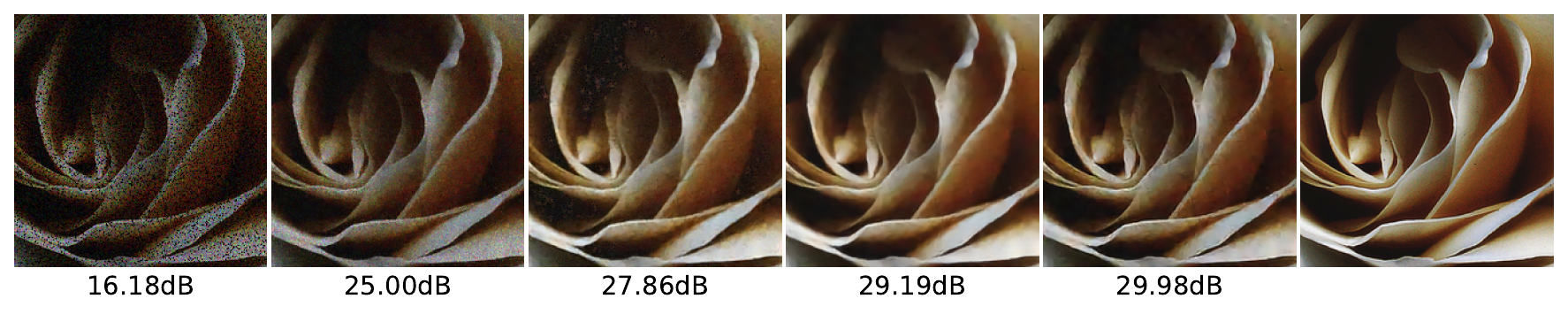}
    \includegraphics[width=\linewidth]{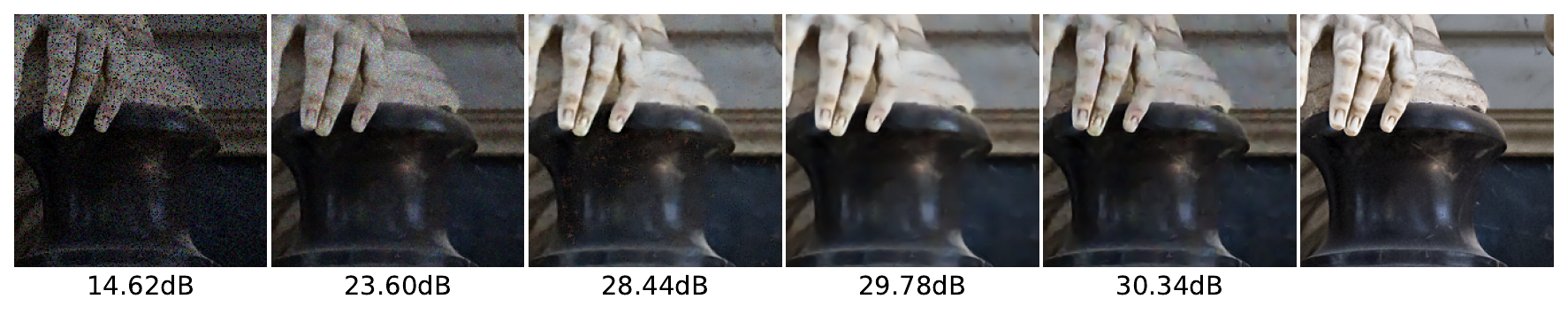}
    \vspace{-2em}\caption{Inpaiting with Poisson noise in DIV2K Dataset.}
    \label{fig:poisson_inpainting_v2k}
\end{figure}

\begin{figure}[!h]
    \centering
    \includegraphics[width=\linewidth]{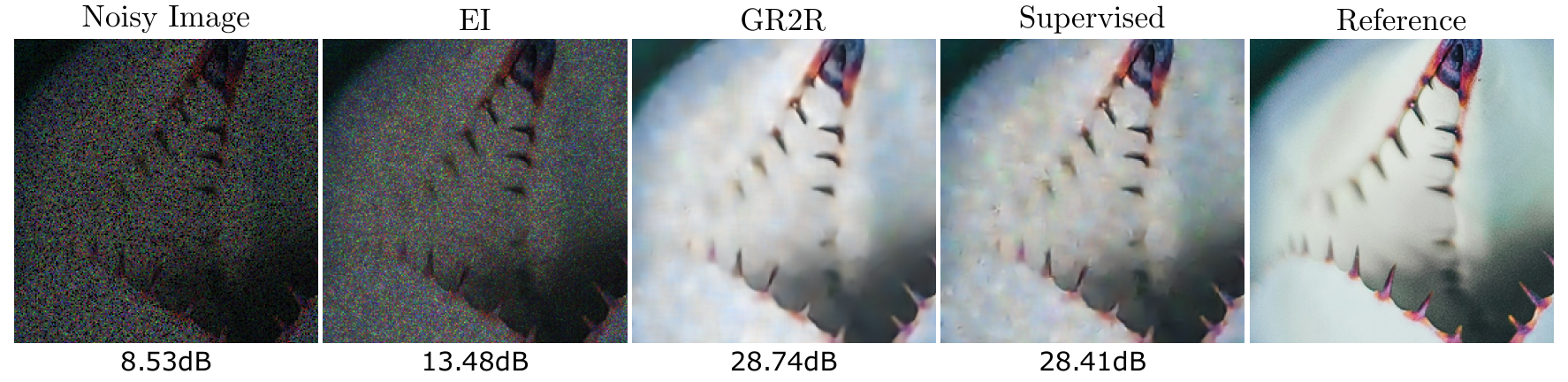}
    \includegraphics[width=\linewidth]{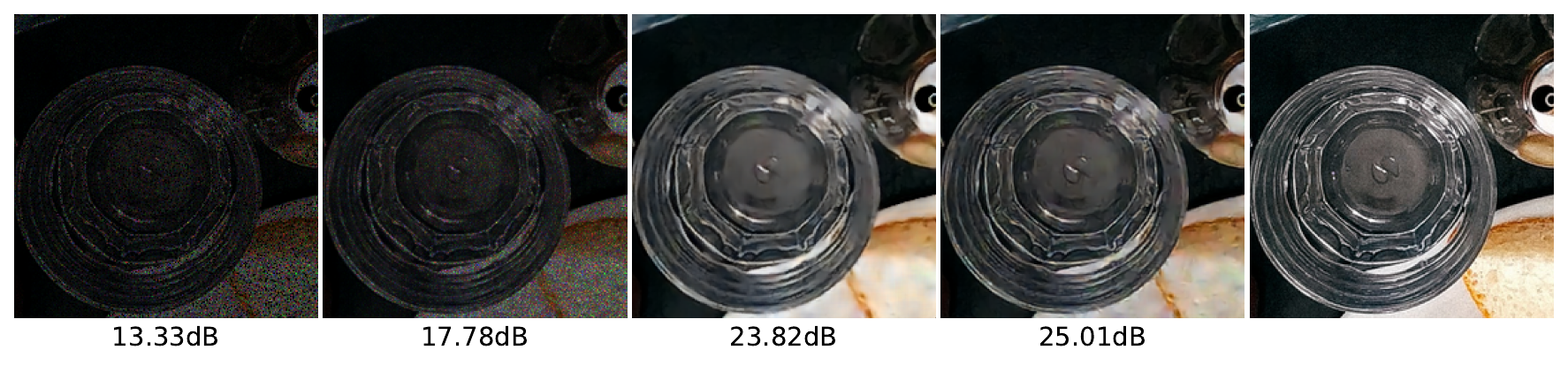}
    \includegraphics[width=\linewidth]{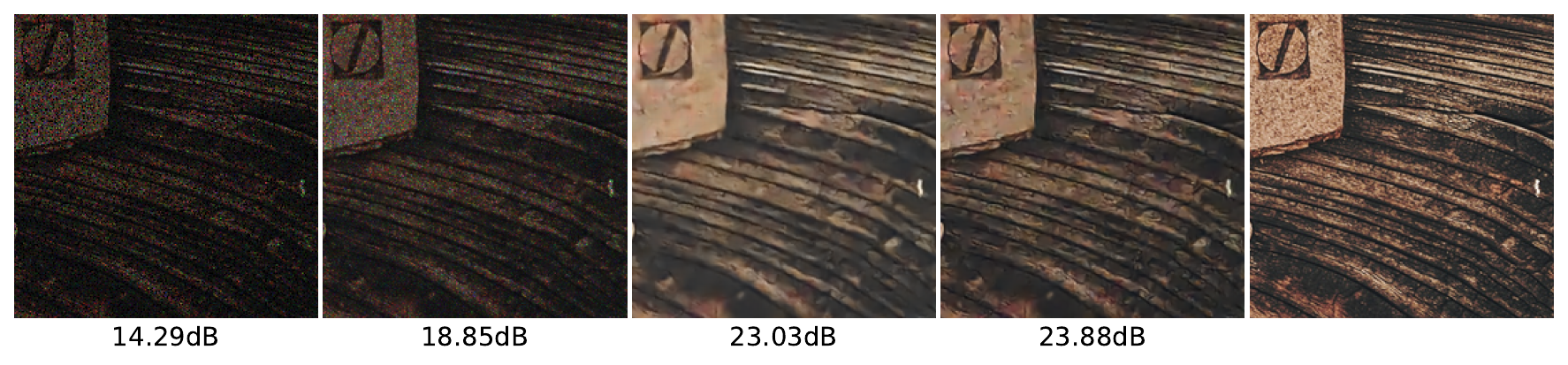}
    \includegraphics[width=\linewidth]{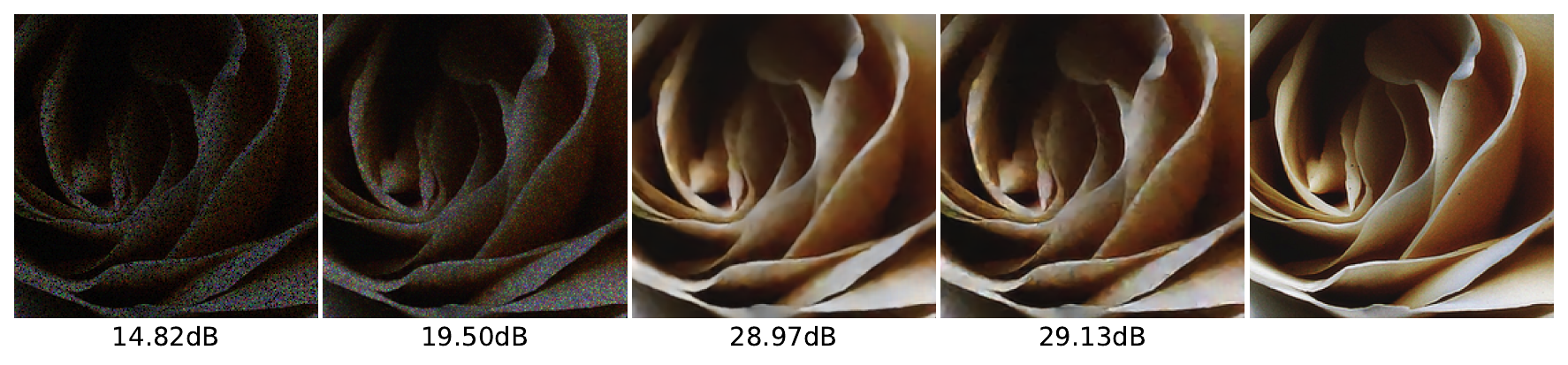}
    \includegraphics[width=\linewidth]{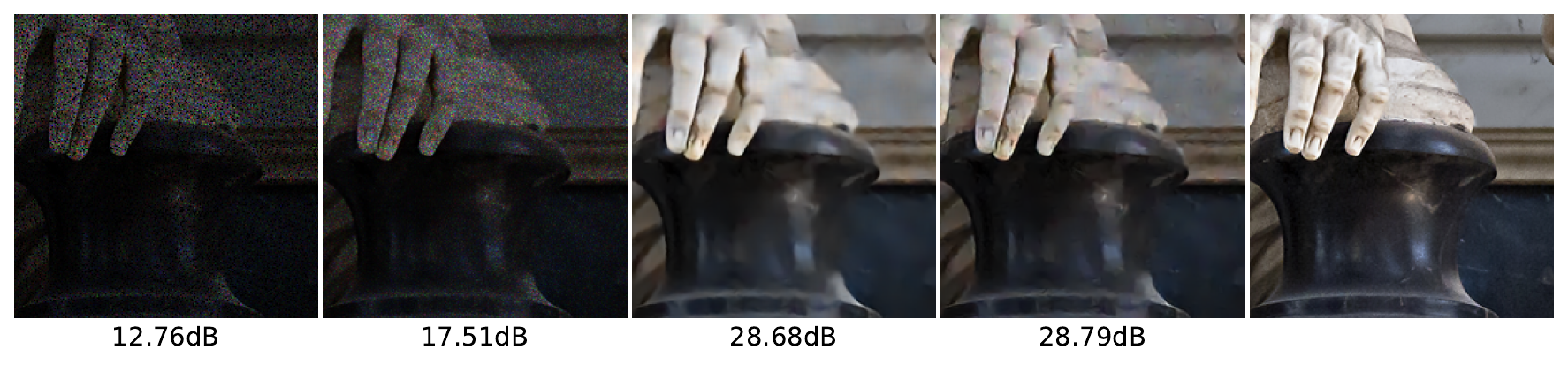}
    \caption{Inpaiting with Gamma noise in DIV2K Dataset.}
    \label{fig:gamma_inpainting_v2k}
\end{figure}

\begin{figure}[!h]
    \centering
    \includegraphics[width=\linewidth]{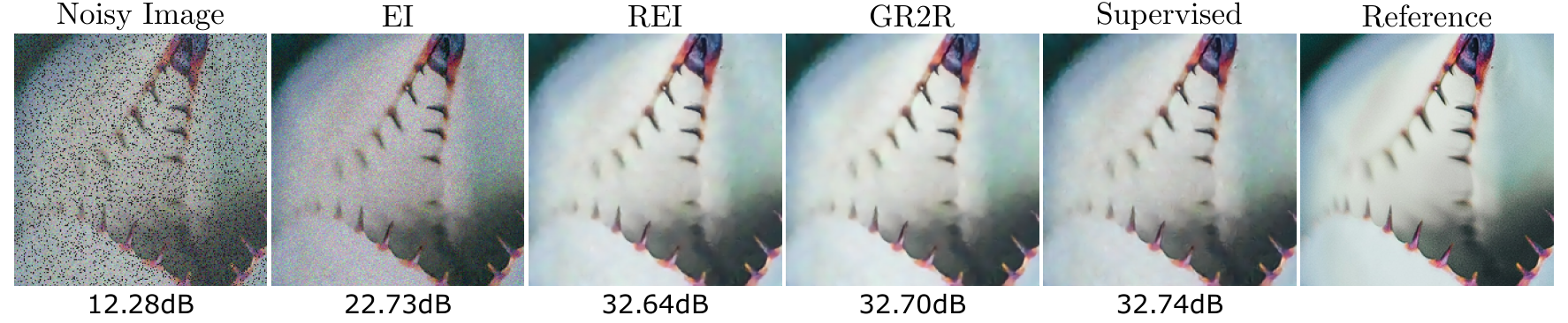}
    \includegraphics[width=\linewidth]{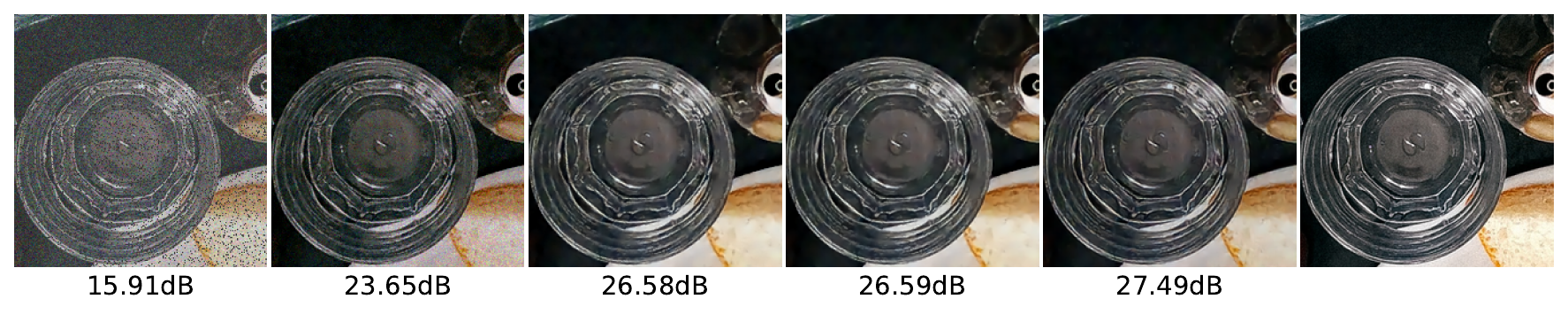}
    \includegraphics[width=\linewidth]{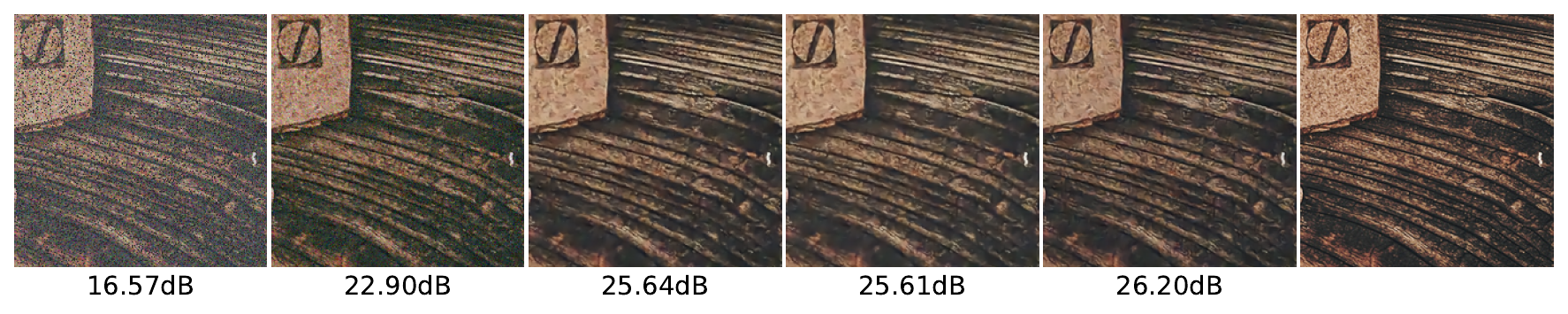}
    \includegraphics[width=\linewidth]{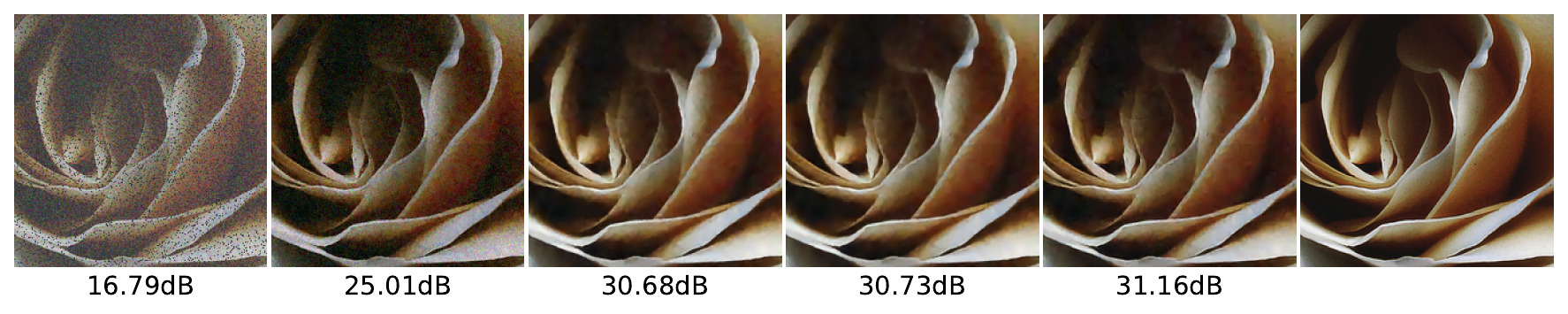}
    \includegraphics[width=\linewidth]{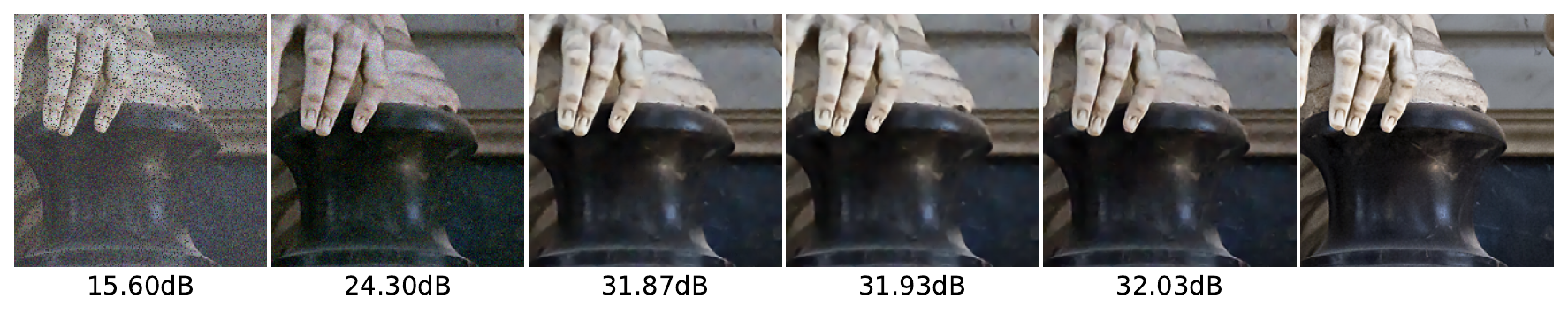}
    \caption{Inpaiting with Gaussian noise in DIV2K Dataset.}
    \label{fig:gaussian_inpainting_v2k}
\end{figure}


\end{document}